# Data-Driven Prediction of CRISPR-Based Transcription Regulation for Programmable Control of Metabolic Flux

Jiayuan Sheng[†], Weihua Guo[†], Christine Ash, Brendan Freitas, Mitchell Paoletti, and Xueyang Feng[*]

Department of Biological Systems Engineering, Virginia Polytechnic Institute and State University, Blacksburg, VA 24061

† WG and JS are equally contributed.
* To whom correspondence should be addressed. X.F: Phone: (540) 231-2974. E-mail: xueyang@vt.edu.

**Abstract.** Multiplex and multi-directional control of metabolic pathways is crucial for metabolic engineering to improve product yield of fuels, chemicals, and pharmaceuticals. To achieve this goal, artificial transcriptional regulators such as CRISPR-based transcription regulators have been developed to specifically activate or repress genes of interest. Here, we found that by deploying guide RNAs to target on DNA sites at different locations of genetic cassettes, we could use just one synthetic CRISPR-based transcriptional regulator to simultaneously activate and repress gene expressions. By using the pairwise datasets of guide RNAs and gene expressions, we developed a data-driven predictive model to rationally design this system for fine-tuning expression of target genes. We demonstrated that this system could achieve programmable control of metabolic fluxes when using yeast to produce versatile chemicals. We anticipate that this master CRISPR-based transcription regulator will be a valuable addition to the synthetic biology toolkit for metabolic engineering, speeding up the "design-build-test" cycle in industrial biomanufacturing as well as generating new biological insights on the fates of eukaryotic cells.

## Introduction

Metabolic engineering has proven to be tremendously important for sustainable production of fuels[1, 2], chemicals[3, 4], and pharmaceuticals[5, 6]. One of the critical steps in metabolic engineering is reprogramming metabolic fluxes in host cells to optimize the fermentation performance such as product yield[7, 8]. To achieve this goal, several enzymes need to be activated while in the meantime others need to be repressed[9]. The multiplex and multi-directional control of enzyme expressions is largely executed through transcriptional regulation[9-12]. Recently, the type-II clustered regularly interspaced short palindromic repeats (CRISPR)/Cas9 from *Streptococcus pyogenes* (*Sp*)[13-15] has been repurposed to be a master transcriptional regulator that could activate or repress multiple genes[16-18]. By further extending the guide RNAs of SpCas9 to include effector protein recruitment sites and expressing the effector proteins in host cells[18, 19], a synthetic CRISPR-based transcriptional regulator was developed to simultaneously program the expressions of multiple genes at multiple directions (i.e., both activation and repression).

Although demonstrating great promises in controlling metabolic fluxes, the current CRISPR-based transcription regulation faces several challenges when being applied for metabolic engineering. First, it relies on a panel of well-characterized genetic parts (e.g., RNA-binding proteins) to achieve the transcriptional regulation[9, 20]. However, the rareness of such genetic parts often limits the application of current CRISPR-based transcription regulator in a metabolic network. Second, it requires the co-expression of effector proteins to activate or repress the target genes[9, 20]. Therefore, the utility of current CRISPR-based transcription regulator in metabolic engineering is mitigated by the metabolic costs associated with protein expressions[9, 20]. To address these limitations, an ideal type of CRISPR-based transcription regulation should meet two criteria: generally applicable in any gene of interest, and requiring minimal protein expression to achieve multi-directional gene regulation.

In this study, we have developed a data-driven approach that enables the rational design of a new type of CRISPR-based transcription regulation. This CRISPR-based transcription regulator uses only a fused protein of a nuclease-deficient Cas9 (dCas9) and an effector (VP64) to achieve multi-directional and multiplex gene regulation, which eliminates the metabolic costs associated with the expression of effector proteins in previous studies[9, 20]. We found that the deployment of guide RNAs was the key factor that determined the regulatory effects of our CRISPR-based transcription regulator. We used a data-driven approach to provide accurate and target-oriented guidance on designing the guide RNAs. As we showed in our results, this approach could be applied for any gene of interest. Finally, using this system, we demonstrated a highly programmable control of metabolic fluxes when using yeast to produce versatile chemicals.

## Results

**Gene activation and repression by using a CRISPR-based transcriptional regulator.** The CRISPR-based transcriptional regulator is composed of a codon-optimized, catalytically dead SpCa9 (i.e., dCas9) that is fused with four tandem copies of Herpes Simplex Viral Protein 16 (VP64, a commonly used eukaryotic transcription activator domain). The similar molecular design was previously reported to be able to activate gene expression in yeast[18] and mammalian cells[18]. In brief, it was found that when the dCas9-VP64 regulator was positioned in the correct sites of promoter, the target gene could be activated by the VP64. However, we hypothesized that the effects of dCas9-VP64 could be diverse, i.e., both activation and repression could be achieved by using this master regulator (Fig. 1A). For example, the transcription initiation could be blocked when dCas9-VP64 is deployed to the transcription starting sites (TSS). The transcription elongation could also be inhibited when dCas9-VP64 is deployed to the open reading frame (ORF). If the hypothesis stands, we could then repurpose dCas9-VP64 as a universal regulator to activate and repress gene expressions at the same time.

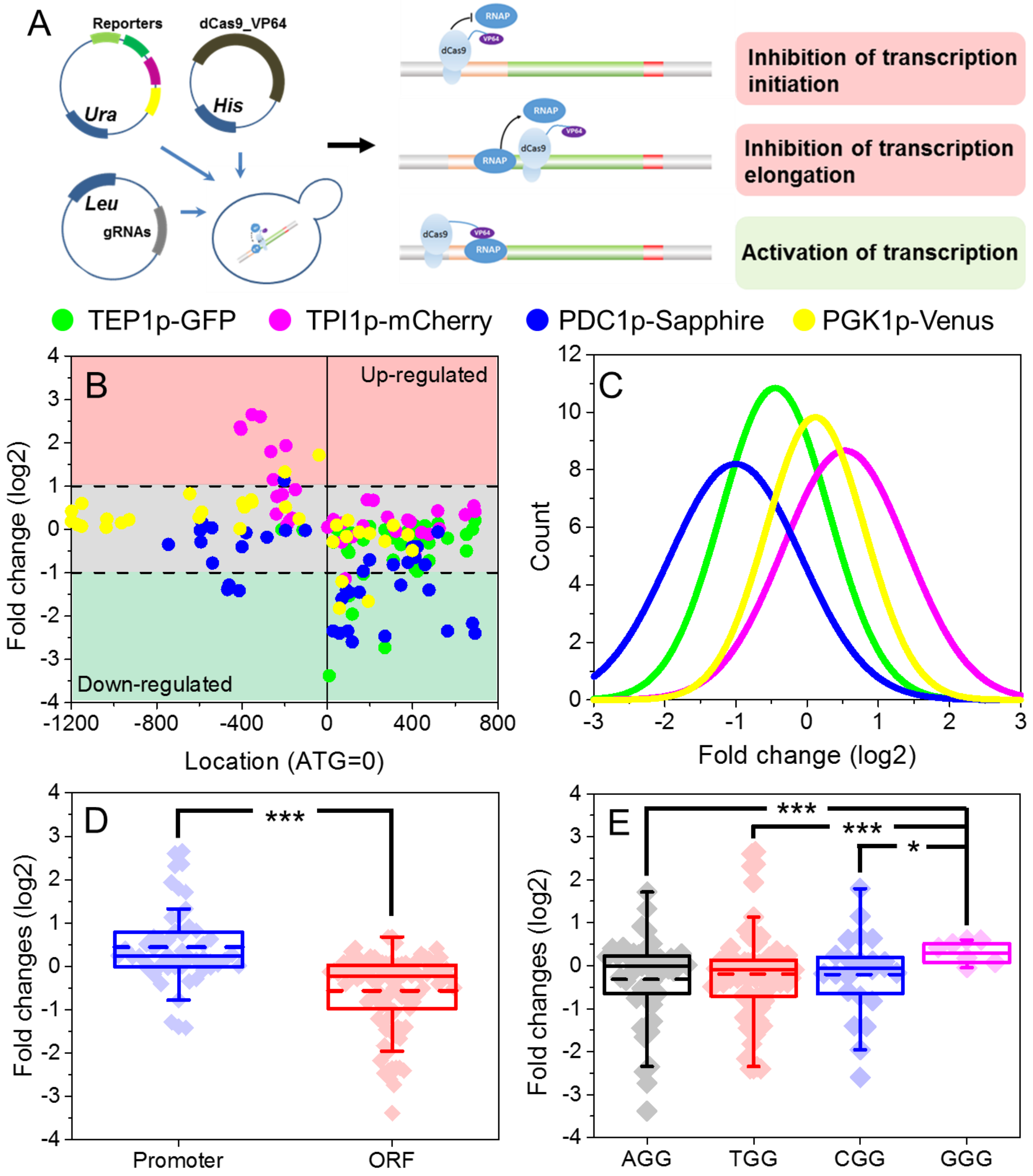

**Figure 1.** Multi-directional transcriptional regulation by dCas9-VP64. (A) Hypothesized mechanism of transcriptional regulation by dCas9-VP64 to achieve both gene activation and gene repression. (B) The measured fold changes of gene expressions from the four synthetic genetic cassettes based on the PAM position. (C) The distributions of fold changes of gene expressions. (D) Comparison of fold changes of gene expressions from two groups: PAM sites located in the promoter regions and PAM sites located in the ORF region. ***: $p < 0.01$. (E) Effects of different PAM sites on transcriptional regulation by dCas9-VP64. *: $p < 0.05$.

To validate our hypothesis, we designed experiments by selecting 138 sites that can be targeted by dCas9-VP64 in four synthetic genetic cassettes (Fig. 1B): GFP under TEF1p promoter, mCherry under TPI1p promoter, Sapphire under PGK1p promoter, and Venus under PDC1p promoter. We created a library of guide RNAs and co-expressed them with the dCas9-VP64 and the target genetic cassette. For each of the tests, we measured the fluorescence of the reporter proteins and compared it to a control test in which dCas9-VP64 and the genetic cassette were expressed without guide RNA. As summarized in Fig. 1B and Fig. 1C, expression of reporter genes could indeed be programmed to be either up- or down-regulated (cutoff fold-change set as two-fold, $p<0.05$) when positioning the guide RNA at different sites on the promoter or ORF. The dynamic range of transcriptional regulation via dCas9-VP64 varied for different genetic elements, with the largest dynamic range achieving 13.8-fold (from -1.14 to 2.65 of log2 gene fold-change) in expression of TPI1p-mCherry cassette and the smallest $\log_2$ fold change dynamic range achieving 11.6-fold (from -1.82 to 1.71 of log2 gene fold-change) in expression of PGK1p-Venus cassette. We next compared the effects on transcriptional regulation when guide RNAs were positioned in the promoter region to that were positioned in the gene region (Fig. 1D). A significant difference ($p<0.01$) was revealed: while gene expression was in general down-regulated when guide RNAs were positioned in the gene region, most of the guide RNAs (94.2%) that were positioned in the promoter region led to gene up-regulation or no effect. We also evaluated whether different PAM sites (i.e., 3'AGG, 3'TGG, 3'CGG, 3'GGG) could bias the transcriptional regulation (Fig. 1E). We found that most of the PAM sites (3'AGG, 3'TGG, 3'CGG) have no bias, but when guide RNAs were targeted on 3'GGG sites, the expression of selected genes tends to be more up-regulated than other PAM sites ($p<0.05$). Such difference, as we discussed in the section below, may be attributed to the larger percentage of guanosine in 3'GGG sites. Finally, we evaluated the metabolic costs of dCas9-VP64 by comparing the growth rates of yeast between the ones subject to transcriptional regulation (i.e., with guide RNAs) and the ones that were not. As shown in Fig. S1, 131 out of the 138 tests showed no significant difference ($p>0.05$) on cell growth rate when being compared to that of the control strain, indicating a minimal metabolic burden when using only one synthetic protein for transcriptional regulation.

**Data-driven model of transcriptional regulation by using dCas9-VP64.** To determine the rule underlying transcriptional regulation by dCas9-VP64, we solicited nine design parameters on nucleotide stability, sequence of the target genetic element, PAM site location, and protein-DNA structure. These design parameters were chosen based on previous studies on the activity of SpCas9 [21, 22]. Next, we correlated these design parameters with transcriptional regulation (Fig. S2A), and calculated the Pearson's correlation coefficients (PCC). The top 3 correlated design parameters were location, GC content (GC%), and PAM site of GGG. Next, we aimed to develop a predictive model that could use the design parameters to describe the effects of guide RNAs positioning on gene expressions. Our first attempt is a linear regression model, which utilized all the nine design parameters (i.e., GC%, location, number of base G, number of base A, $\Delta G$, AGG, TGG, CGG, and GGG) to simulate the corresponding fold changes of gene expressions as we collected from the four synthetic genetic cassettes (Fig. S2B). However, the fitting of the linear regression model was very bad, as demonstrated by the low PCC (0.41) between observed and simulated fold changes (Fig. S2B). This clearly indicated that the biomolecular interactions of RNA-protein and DNA-protein in the dCas9-VP64 system were highly nonlinear, which cannot be captured by the simple linear regression model.

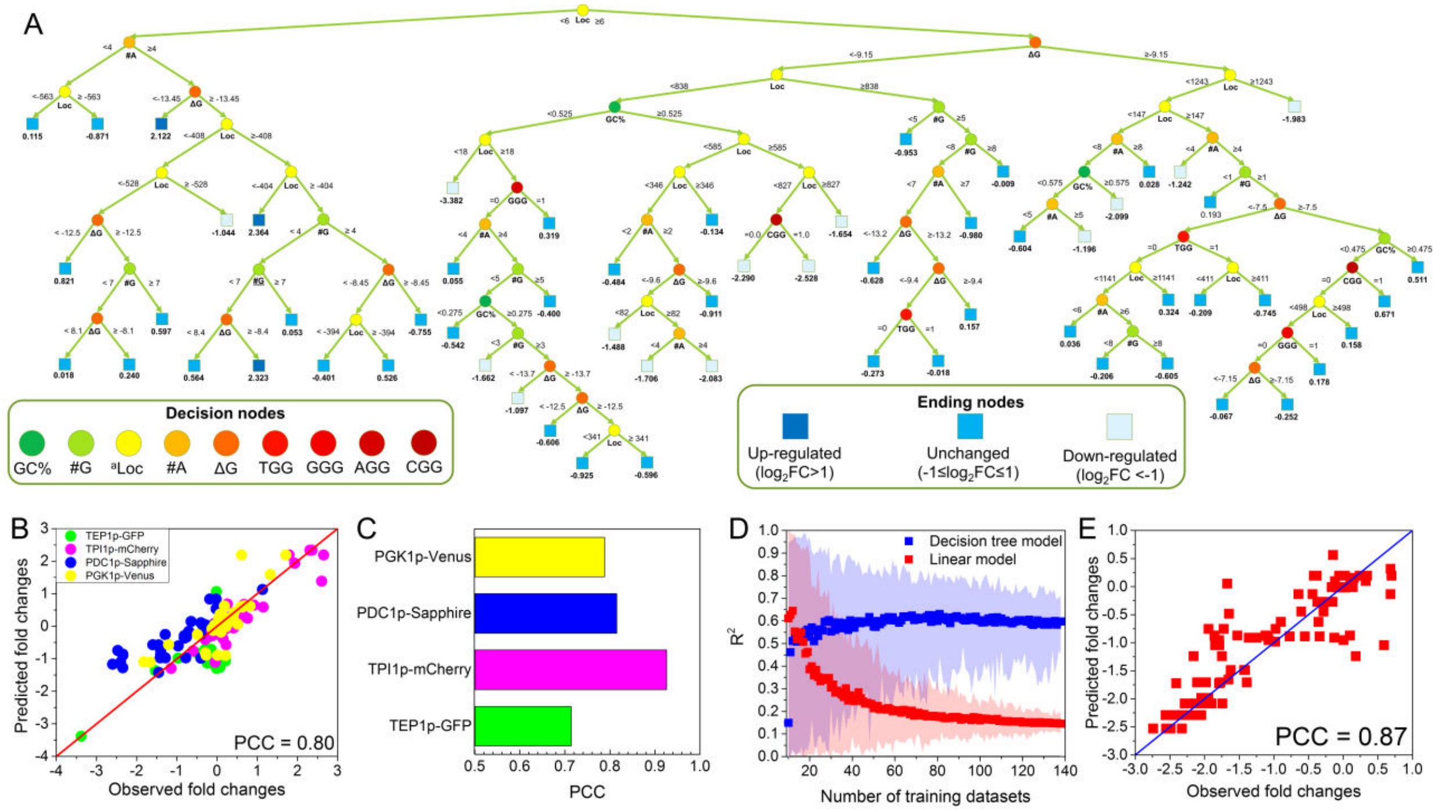

**Figure 2.** Data-driven model of transcriptional regulation by using dCas9-VP64. (A) Binary regression tree model trained with all the screening data from the four synthetic genetic cassettes. The regression tree model consisted of 58 decision nodes and used six design parameters of guide RNAs as input. (B) and (C) Prediction accuracy of the regression tree model from ten-fold cross validation. (D) Impact of data size on model prediction. For data-driven model, the prediction increased with the inclusion of more datasets. For linear model, the prediction decreased when more datasets were used. The shadow areas indicate the 95% confidence interval of model prediction. (E) Validation of the regression tree model by comparing the simulated and experimentally measured gene regulations on Eno2p-tdTomato cassette subject to dCas9-VP64 regulation.

We then used machine learning to derive an empirical model to quantitatively predict the nonlinear correlation between design parameters and transcriptional regulation by dCas9-VP64. As a data-driven modeling approach, machine learning is advantageous than linear model in solving complex problems in two aspects[23-25]: requiring no *a priori* knowledge of the system, and capable of resolving complex systems with high non-linearity and multi-dimensionality. In this study, we used the pairwise data of the design parameters of guide RNAs and the corresponding fold-change of gene expressions to train the computer for developing a mathematical model that could accurately predict the causal effects of inputs (i.e., regulated gene expression in this study). We used decision tree method to build a machine-learning algorithm and adopted ten-fold cross validation to evaluate the prediction accuracy of our model (Fig. 2A). The model construction and model evaluation were conducted by following a toolkit developed in MATLAB™ (i.e., "*fitrtree*" and "*predict*" in Statistics and Machine Learning Toolbox), which automatically adjust the nodes and connections of the decision tree to optimize the fitting[26, 27]. Using the 138 pairwise data collected from the synthetic genetic cassettes (i.e., TEF1p-GFP, TPI1p-mCherry, PDC1p-Sapphire, and PGK1p-Venus), we found that the prediction accuracy was dramatically improved with PCC reaching 0.80 (Fig. 2B) for overall prediction of gene regulation and 0.72~0.93 for predicting gene regulation of individual genetic cassettes (Fig. 2C). Also, we found that the prediction accuracy of our machine-learning algorithm was improved with the enlargement of data size (Fig. 2D). For example, when 20 pairwise datasets were chosen, the PCC was merely 0.66. However, when the number of pairwise datasets reached 80, the PCC was improved to 0.78. This demonstrated the unique advantage of data-driven algorithm, i.e., increased prediction accuracy with more data.

To further test if our machine-learning algorithm could be generally applied for predicting the transcriptional regulation of other genes, we designed another synthetic genetic cassette that expressed tdTomato under Eno2p promoter. We used the machine-learning algorithm to predict the regulated gene expressions when guide RNAs of dCas9-VP64 were positioned at different locations of Eno2p-tdTomate cassette, followed by constructing the genetic cassettes and conducting experimental measurements. Of the 99 guide RNAs tested, our machine-learning algorithm could achieve similarly high prediction accuracy with PCC between the predicted and the measured gene expressions reaching 0.87 (Fig. 2E). This success demonstrated that our data-driven model could be generally applicable to guide the biomolecular design of dCas9-VP64 system to achieve customized gene regulation.

We also packaged our machine-learning algorithm into an open-source toolbox (Fig. S3 and supplementary software), CRISTINES (CRISPR-Cas9 Transcriptional Inactivation aNd Elevation System), which is a MATLAB-based toolbox and free for downloading at https://sites.google.com/a/vt.edu/biomolecular-engineering-lab/software. CRISTINES is able to analyze the input DNA sequences, identify the design parameters of the dCas9-VP64 system, and use the embedded decision-tree algorithm to provide the top five guide RNAs that would lead to the strongest up-regulation and down-regulations, respectively. We anticipated that CRISTINES could help biologists worldwide to customize their design based on the target gene of interest.

**Design dCas9-VP64 to reprogram metabolic fluxes in yeast.** We next applied dCas9-VP64 system in yeast metabolic engineering to test if the metabolic fluxes could be flexibly reprogrammed using our CRISPR-based transcription regulator. We chose the highly branched bacterial violacein biosynthetic pathway as our model pathway[28] (Fig. 3A), which uses five enzymes (VioA, VioB, VioE, VioD, and VioC) to produce four high-value products (violacein, proviolacein, deoxyviolacein, and prodeoxyviolacein). By controlling the expression levels of the five enzymes, the metabolic fluxes flowing into different branch pathways could be varied and thus leading to different yield of the four products. In this study, we reconstituted the violacein pathway in yeast by expressing the five enzymes under constitutive promoters (i.e., TEF1p, PGK1p, ENO2p, TPI1p, and PDC1p). According to our data-driven model CRISTINES, we could computationally predict the gene expression and thus predict the metabolic fluxes in the violacein pathway. Here, we designed three guide RNAs for four genes in the violacein pathway (VioA, VioE, VioD, and VioC). These three guide RNAs targeted on different promoter sites and were predicted to result in high, medium and low expression of each target gene, respectively. Correspondingly, the functional output states of the violacein pathway were predicted to vary. For example, up-regulating VioA and down-regulating VioD would increase the fluxes into deoxyviolacein and prodeoxyviolacein, but decrease the fluxes into violacein and proviolacein. To validate our predictions on metabolic flux reprogramming, we co-expressed the dCas9-VP64 system with the violacein pathway, and used various guide RNAs to fine tune gene expressions. For each of the tests, we measured the titer of violacein, proviolacein, deoxyviolacein, and prodeoxyviolacein produced by yeast. As shown in Fig. 3B and 3C, our prediction fit well with the experimental measurements, with PCC reaching 0.84. We noticed some of the predictions on metabolic fluxes were not as good as we expected. This could be attributed to the posttranscriptional regulation of the five enzymes in the violacein pathway. Overall, we demonstrated that our master CRISPR-based transcription regulator was indeed able to program metabolic fluxes. More importantly, our data-driven algorithm allows users to design metabolic pathways with deterministic fates *in silico*.

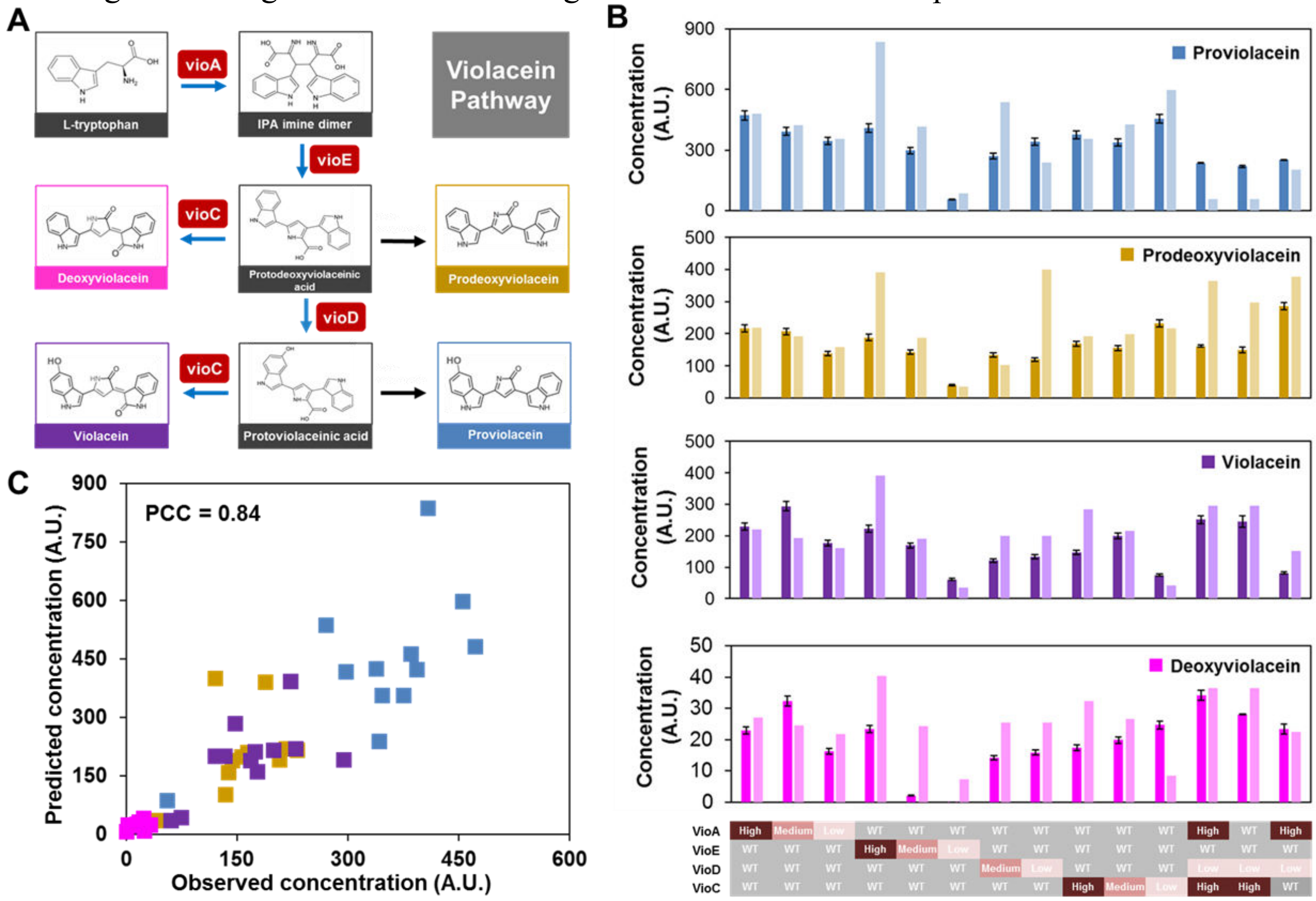

**Figure 3.** Design dCas9-VP64 to reprogram metabolic fluxes in yeast. (A) Violacein pathway in yeast used in this study to demonstrate the programmable control of metabolic fluxes by using dCas9-VP64. Five enzymatic steps (vioA, vioB, vioC, vioD, and vioE) and two non-enzymatic steps led to four products from the violacein pathway: proviolacein, prodeoxyviolacein, violacein, and deoxyviolacein. (B) A panel of genes subject to regulation of dCas9-VP64 were chosen to control metabolic fluxes to various products from the violacein pathway. WT: wild type gene without any regulation; High: highly up-regulated gene expression by dCas9-VP64; Medium: medium-level up-regulated gene expression by dCas9-VP64; Low: down-regulated gene expression by dCas9-VP64. (C) The compassion between the model-predicted and experimentally measured products from the violacein pathway. The high correlation coefficient (PCC=0.84) indicated that the data-driven model could accurately predict the effects of artificial Cas9-based regulator on metabolic flux reprogramming.

## Discussion

Using data-driven approach to investigate biomolecular interactions of CRISPR-based systems has recently been showcased in several studies, such as rational design of guide RNAs for maximizing editing activity and minimizing off-target effects of SpCas9[29, 30]. This approach is advantageous compared to conventional deterministic models because it does not require *a priori* knowledge on the mechanisms of RNA-protein and DNA-protein interactions[31, 32], which still remains largely unknown in spite of numerous studies on SpCas9 structures[33-37]. Also, because the biomolecular interactions among Cas9 protein and nucleic acids are highly nonlinear, the linear regression model cannot capture the essence of CRISPR-based transcription regulation. As shown in our results, a machine-learning method could overcome this issue and capture the nonlinearity of the model. Not only did we achieve high accuracy when predicting the regulatory effects of CRISPR-based transcription regulator, but we also demonstrated that this method was not specific to a few selected genes and could be generally applicable. Future work will determine if the results obtained from yeast could provide useful lessons in other eukaryotic systems such as mammalian cells.

We expect that our method will provide a valuable tool for metabolic engineering, especially yeast metabolic engineering at this stage. *S. cerevisiae* is a widely used industrial workhorse for producing a broad spectrum of chemicals that represents over quarter trillion dollars market[38]. The experimental and analytical approaches described here raise the possibility of genome-scale reprogramming of metabolic fluxes, which will dramatically speed up the "design-build-test" cycle in industrial biomanufacturing[39]. We also expect our method could be used to rewire the fate of yeast cells, such as cell cycle, and thus generate new biological insights on the fundamentals of metabolic diseases, aging and apoptosis by using yeast as a disease model.

## Methods

**Strain and plasmid construction in *Saccharomyces cerevisiae*.** dCas9 was codon-optimized for expression in *S. cerevisiae* and cloned into a pRS413 backbone under control of the GAL1 promoter. The RNA-guided transcription factors were built by fusing four repeats of the minimal domain of the herpes simplex viral protein 16 (VP16) to the C-terminus of dCas9 (dCas9_VP64)[18]. The reporter genes eGFP under the control of the TEF1p promoter, sapphire under PDC1p, mCherry under TPI1p, and venus under PGK1p were cloned into pRS416 plasmid by using the DNA assembler method[40, 41]. The reporter plasmid for verification was built by cloning tdTomato under the control of ENO2p into pRS416 plasmid. To build gRNA-expressing plasmids, empty gRNA expressing vectors were first made by cloning the pRPR1 promoter (an RNA-polymerase-III-dependent promoter), the gRNA handle (flanked by HindIII and Xho1 sites), and the RPR1 terminator into the SacI and KpnI sites of the pRS425 plasmid. Sequences of the constructs that were used in this study were listed in Table S1. Strains constructed in this study were listed in Table S2.

**Fluorescence Assays**. To assess expression of the reporter constructs, yeast cells expressing different gRNAs (or no gRNA as control) were grown overnight (250 rpm, 30°C) in 3 mL SC medium supplemented with glucose with appropriate selection (three independent cultures for each sample). Ten microliters of these cultures were then transferred into fresh media, supplemented with galactose and grown for 20 h (250 rpm, 30°C) before analysis by plate reader. The wave lengths of the different reporter genes are eGFP: $\lambda_{ex}$ 488 nm, $\lambda_{em}$ 507nm; Sapphire: $\lambda_{ex}$ 399 nm, $\lambda_{em}$ 511nm; Venus $\lambda_{ex}$ 515 nm, $\lambda_{em}$ 528nm; tdTomato $\lambda_{ex}$ 554 nm, $\lambda_{em}$ 581nm; mCherry $\lambda_{ex}$ 587 nm, $\lambda_{em}$ 610 nm. All of the fold-changes of the synthetic genetic cassettes, including the four cassettes for model training and the ENO2p-tdTomato cassette for model validation, were listed in Table S3 and Table S4.

**Qualitative analysis of key parameters.** The exponential fold changes of different gRNA designs have been separated into two categories, the promoter region (location < 0, n = 52) and the gene coding region (location > 0, n = 86). The unpaired two-tail t-test was used to calculate the significance of the fold changes between these two groups. For the analysis of the PAM type, the same t-test was used for each PAM types ($n_{AGG} = 42$, $n_{TGG} = 69$, $n_{CGG} = 21$, $n_{GGG} = 6$).

**Modeling.** The multiple linear regression model was implemented by the "*regress*" command in MATLAB. To evaluate the prediction power, the ten-fold cross validation was implemented for the linear model. The Pearson's correlation coefficient between observed and predicted fold changes was calculated by MATLAB. The binary regression decision tree model was developed by using "*fitrtree*" command in MATLAB with all the default setting of options. The details of the decision tree model were shown in Table S5. The ten-fold cross validation was implemented for this model with the same manner. A detailed tutorial of CRISTINES was included in the supplementary information.

**Ten-fold cross validation.** All the datasets from the four synthetic genetic cassettes (n = 151) were randomly divided into ten folds (nine folds with 15 datasets each and one-fold with 16 datasets). Nine of the ten folds were used as training datasets, and the one-fold remaining was used as validation datasets. By iteratively repeating the above process for ten times, all the folds could be used as validation datasets. The Pearson's correlation coefficient between the observed fold changes and predicted fold changes of were calculated by MATLAB.

**Analysis of products from violacein pathway.** Yeast strains for violacein biosynthesis were constructed and product distributions were analyzed as described previously with minor modifications. The parent yeast strain for these experiments was BY4741. The five-gene cassette of violacein pathway was constructed using the DNA assembler method: VioA under TEF1p; VioB under PGK1p; VioC under ENO2p; VioD under TPI1p and VioE under PDC1p. Yeast strains with violacein pathway genes and the CRISPR system with constitutive dCas9 expression were grown in SC medium containing 5% galactose. After 3 days at 30 °C, approximately 2 mL of yeast cultures were harvested and the cells were collected and suspended in 250 µL of methanol, boiled at 95 °C for 15 minutes, and vortexed twice during the incubation. Solutions were centrifuged twice to remove cell debris, and the products from violacein pathway (i.e., violacein, proviolacein, deoxyviolacein, and prodeoxyviolacein) in the supernatant were analyzed by HPLC on an Agilent Rapid Resolution SB-C18 column as described previously, measuring absorbance at 565 nm[42].

**Code availability.** We claim that the code mentioned in this study is available within the article's Supplementary Information files (Supplementary Software) and at https://sites.google.com/a/vt.edu/biomolecular-engineering-lab/ as a MATLAB package file (MathWorks Inc.).

**Data availability.** All data generated or analyzed during this study are included in this published article and its Supplementary Information files.

**Acknowledgement**
We thank the Writing Center in Virginia Tech for improving the language of the paper. This study was supported by start-up fund (#175323) and the ICTAS Junior Faculty Award from Virginia Tech.

**Author contributions**
X.F. contributed to the initial idea of this project. J.S., C.A., and B.F. contributed to the molecular biology experiments and data screening. W.G. and X.F. contributed to the computational modeling and analysis. J.S. and W.G. contributed to the experimental validation and violacein pathway showcase. W.G. and M.P. contributed to the offline software development. J.S., W.G., and X.F. contributed to the manuscript preparation and revision.

**Competing financial interests.**
The authors do not have any conflicts of financial interests.

**Supplementary Information**
**Figure S1.** The $OD_{600}$ fold change for all the strains used for screening. (A) $OD_{600}$ fold-change for GFP screening set. (B) $OD_{600}$ fold-change for Sapphire screening set. (C) $OD_{600}$ fold-change for mCherry screening set. (D) $OD_{600}$ fold-change for Venus screening set. The results showed that no significant metabolic burden could be detected compared with wild type strain. The strains marked with (*) indicated a significant inhibition of growth ($p<0.05$).
**Figure S2.** Linear model for predicting transcriptional regulation by dCas9-VP64. (A) Pearson's correlation coefficients between each guide RNA design parameters and the fold changes of gene expressions from the screening

data of the four synthetic genetic cassettes. (B) Simulation accuracy of the linear model. (C) Validation of the linear model by comparing the simulated and experimentally measured gene regulations on Eno2p-tdTomato cassette subject to dCas9-VP64 regulation.

**Figure S3.** Screenshot of CRISTINES software with a demo sequence.

**Table S1.** DNA sequences and plasmids used in this study.

**Table S2.** Strains used in this study.

**Table S3.** Key parameters of guide RNAs used in this study.

**Table S4.** Key parameters of guide RNAs used in validation experiments (Eno2p-tdTomato).

**Table S5.** Binary regression decision tree model.

**Table S6.** Measured and predicted concentrations of products from violacein pathway

**Supplementary Software.** CRISTINES toolbox.

**Supplementary method**

## 1. Binary regression decision tree model

### 1.1. Parameterization of guide RNAs

Nine key design parameters, i.e., location, GC content (GC%), minimal free energy (ΔG), number of base A, number of base G, AGG, TGG, CGG, and GGG, have been extracted from guide RNAs, respectively. Location is defined as the number of bases between the first base of guide RNA binding sequence and the A in the start codon (ATG). Therefore, the position of A in the ATG is considered as 0. The location of guide RNAs binding before the start codon is defined as negative, and vice versa. The GC content (GC%) is defined as the proportion of base G and base C in the guide RNA sequence. Minimal free energy (ΔG) is calculated by *rnafold* command in MATLAB. As a categorical parameter, AGG, TGG, CGG, and GGG are four bool parameters representing four types of the PAM site type. Each guide RNA has a set of key design parameters. The parameters are used to train and validate the decision tree model.

### 1.2. Setup of regression decision tree model

To train the binary regression decision tree model, "*fitrtree*" command in MATLAB was used with the default settings. In detail, the standard CART algorithm was used to select the best split predictor, which maximized the split-criterion gain over all possible splits of all predictors. The "*fitrtree*" grew the regression tree and estimated the optimal sequence of pruned subtrees based on the equally weighted mean squared errors, but did not prune the regression tree. During the tree growing, splitting nodes stopped when quadratic error per node dropped below $10^{-6}$ of the quadratic error for the entire data computed before the decision tree was grown. To control the depth of the tree and avoid the over-fitting of the tree model, the maximum splits number was the number of guide RNAs in training datasets.

### 1.3. Application of regression decision tree model

To apply our binary regression decision tree model, "*predict*" command in MATLAB was used to predict the fold changes of gene expression levels for specific guide RNAs. (https://www.mathworks.com/help/stats/compactregressiontree.predict.html?searchHighlight=predict&s_tid=doc_srchtitle ). In general, the target guide RNA sequence will be first processed by customized functions (supplementary software) to calculate the corresponding design parameters. An input matrix will be generated for the tree model, which will be used to predict the corresponding fold change of the gene expression level.

### 1.4 CRISTINES software and its results

CRISTINES software is an offline MATLAB-based tool to provide the guide RNA designs to achieve gene up-regulation and down-regulation. The core algorithm is the binary regression decision tree model that uses the key design parameters of guide RNAs as the input. In the CRISTINES, the sequence of interest with the identification of the start codon ATG, is used as the source file. All the possible guide RNAs from the input sequence will be generated, and the nine design parameters, i.e., location, GC content, minimal free energy, number of base A, number of base G, AGG, TGG, CGG, and GGG, are calculated for each of guide RNAs. These design parameters will then be used in the binary regression decision tree model to calculate the regulatory effects (i.e., fold change of gene expression level) of guide RNAs. CRISTINES will report the top 5 guide RNA designs with highest up- and down- regulation effects, respectively. The regulatory effects of all possible guide RNAs can be downloaded to the local folder in Excel file.

## 2. Calculation of metabolic fluxes in violacein pathway subject to dCas9-VP64 regulation

We hypothesized that the changes gene expression levels would quantitatively change the metabolic fluxes in the branches of violacein pathway, and thus, leading to different concentrations of various products. In detail, the flows through pathways governed by gene *vioA*, *vioB*, *vioC*, *vioD*, and *vioE*, are defined as $f_A$, $f_B$, $f_C$, $f_D$, and $f_E$, respectively. The predicted fold changes of gene *vioA*, *vioB*, *vioC*, *vioD*, and *vioE*, are defined as $FC_A$, $FC_B$, $FC_C$, $FC_D$, and $FC_E$, respectively. Similarly, the flows through the non-enzymatic pathways for producing the final colorful chemical, proviolacein, prodeoxyviolacein, violacein, and deoxyviolacein, are defined as $f_{PV}$, $f_{PDV}$, $f_V$, and $f_{DV}$, respectively. To calculate the effects of dCas9-VP64 on metabolic fluxes in violacein pathway, we assume a pseudo-steady state was achieved. Thus, based on the flux balance at steady state, the quantitative relations among these parameters were shown below:

$$f_A = f_B = f_E \qquad \text{Eq. 1}$$

$$f_E = f_{DV} + f_{PDV} + f_D \qquad \text{Eq. 2}$$

$$f_D = f_V + f_{PV} \qquad \text{Eq. 3}$$

$$f_C = f_{DV} + f_V \qquad \text{Eq. 4}$$

In addition, we hypothesized that the proportion between $f_{DV}$ and $f_V$ obeys the Hill equation as shown in Eq. 5 with Hill coefficient defined as *n*:

$$f_{DV} / f_V = 1 / (f_C^{\,n} + 1) \qquad \text{Eq. 5}$$

The values of $f_{PV}$, $f_{PDV}$, $f_V$, and $f_{DV}$ can be directly obtained from the HPLC measurements, which would serve as the experimental measurements. The constant *n* was calculated as 0.356 from the measurements of the control strain by using Eq. 4 and Eq. 5. The following methods were used to calculate the simulated $f_{PV}$, $f_{PDV}$, $f_V$, and $f_{DV}$.

1) If the dCas9-VP64 system regulates the gene expression of *vioA* or *vioE*, then $f_E = FC_A * f_{E,\text{Control}}$ or $f_E = FC_E * f_{E,\text{Control}}$, respectively; $f_D = FC_D * f_{D,\text{Control}}$, and $f_C = FC_C * f_{C,\text{Control}}$.
2) If the dCas9-VP64 system regulates the gene expression of *vioC*, then $f_C = FC_C * f_{C,\text{Control}}$, $f_E = f_{E,\text{Control}}$, and $f_D = f_{D,\text{Control}}$.
3) If the dCas9-VP64 system regulates the gene expression of *vioD*, then $f_D = FC_D * f_{D,\text{Control}}$, $f_E = f_{E,\text{Control}}$, and $f_C = f_{C,\text{Control}}$.
4) If the dCas9-VP64 system is regulating multiple gene expression in the violacein pathways, we assume that the regulation of each gene is independent to others. In other words, we hypothesized that the regulation of dCas9-VP64 system was independent on each gene and each pathway of the violacein pathways.

By knowing $f_{E}$, $f_{D}$, and $f_{C}$, we could then solve Eq 1-5 together to derive the simulated $f_{DV}$, $f_{V}$, $f_{PV}$, and $f_{PDV}$ sequentially. We next compared the experimentally measured and simulated $f_{DV}$, $f_{V}$, $f_{PV}$, and $f_{PDV}$ as shown in Fig. 3B and 3C.

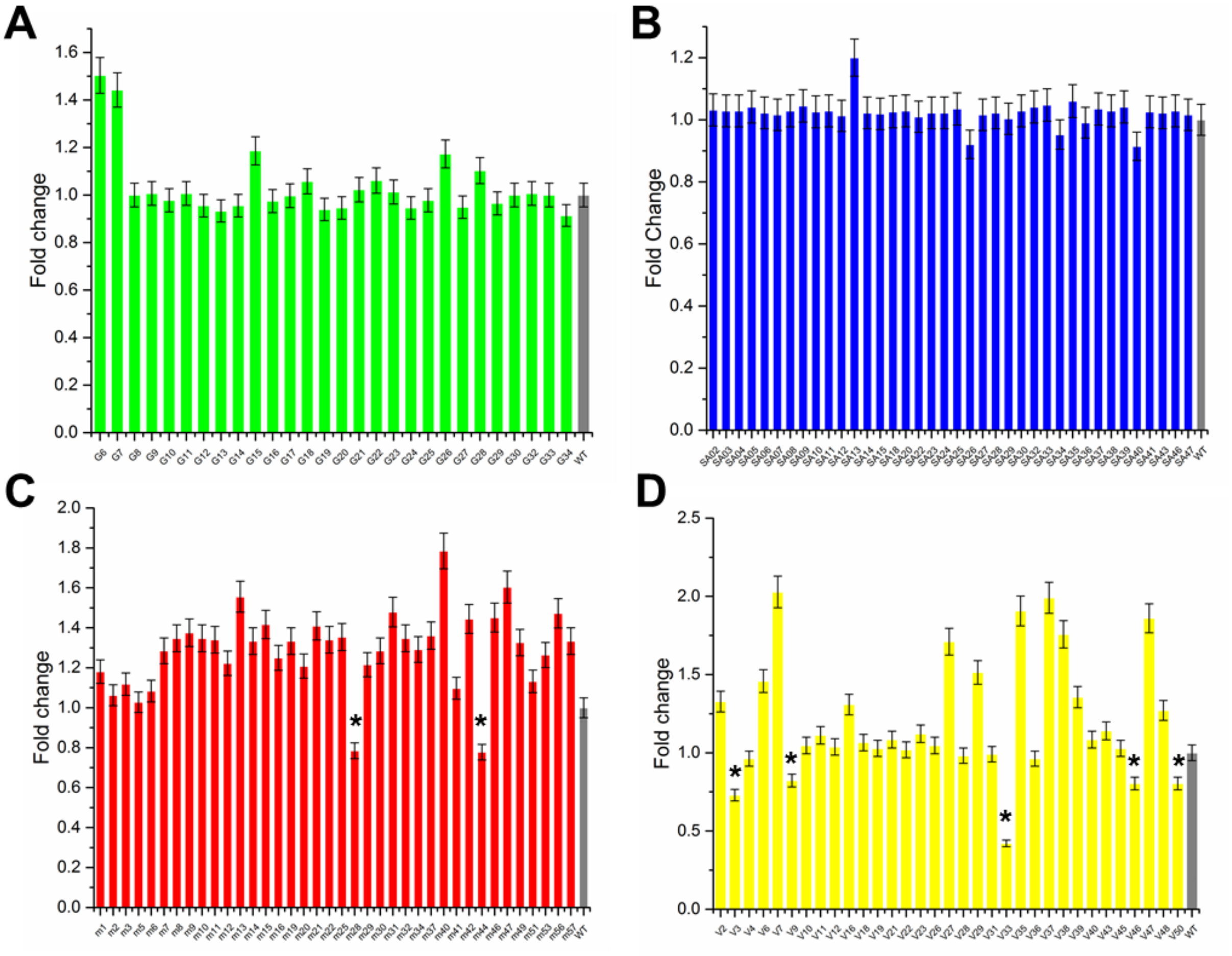

**Figure S1.** The $OD_{600}$ fold change for all the strains used for screening. (A) $OD_{600}$ fold-change for GFP screening set. (B) $OD_{600}$ fold-change for Sapphire screening set. (C) $OD_{600}$ fold-change for mCherry screening set. (D) $OD_{600}$ fold-change for Venus screening set. The results showed that no significant metabolic burden could be detected compared with wild type strain. The strains marked with (*) indicated a significant inhibition of growth ($p$<0.05).

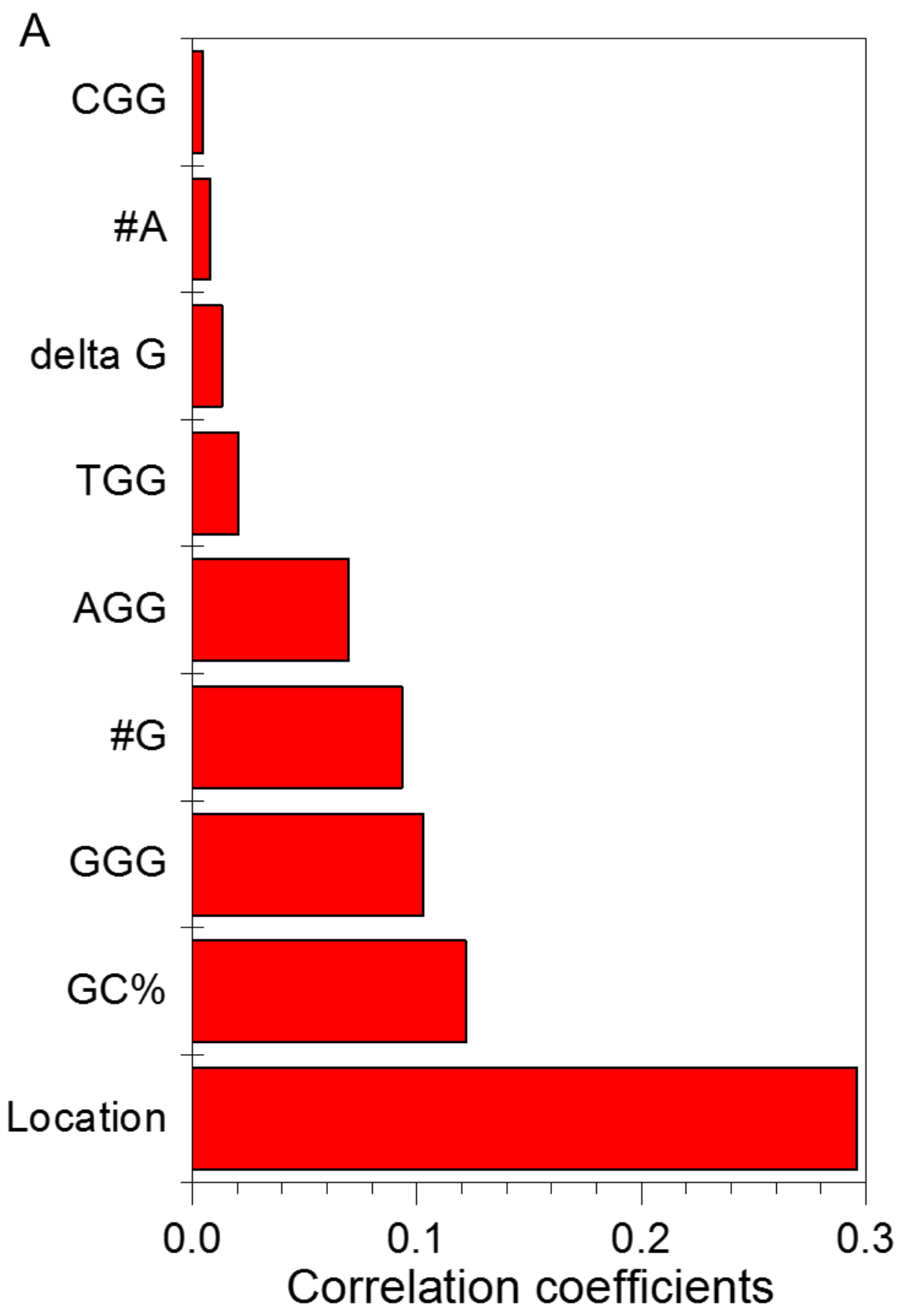

**Figure S2.** Linear model for predicting transcriptional regulation by dCas9-VP64. (A) Pearson's correlation coefficients between each guide RNA design parameters and the fold changes of gene expressions from the screening data of the four synthetic genetic cassettes. (B) Simulation accuracy of the linear model. (C) Validation of the linear model by comparing the simulated and experimentally measured gene regulations on Eno2p-tdTomato cassette subject to dCas9-VP64 regulation.

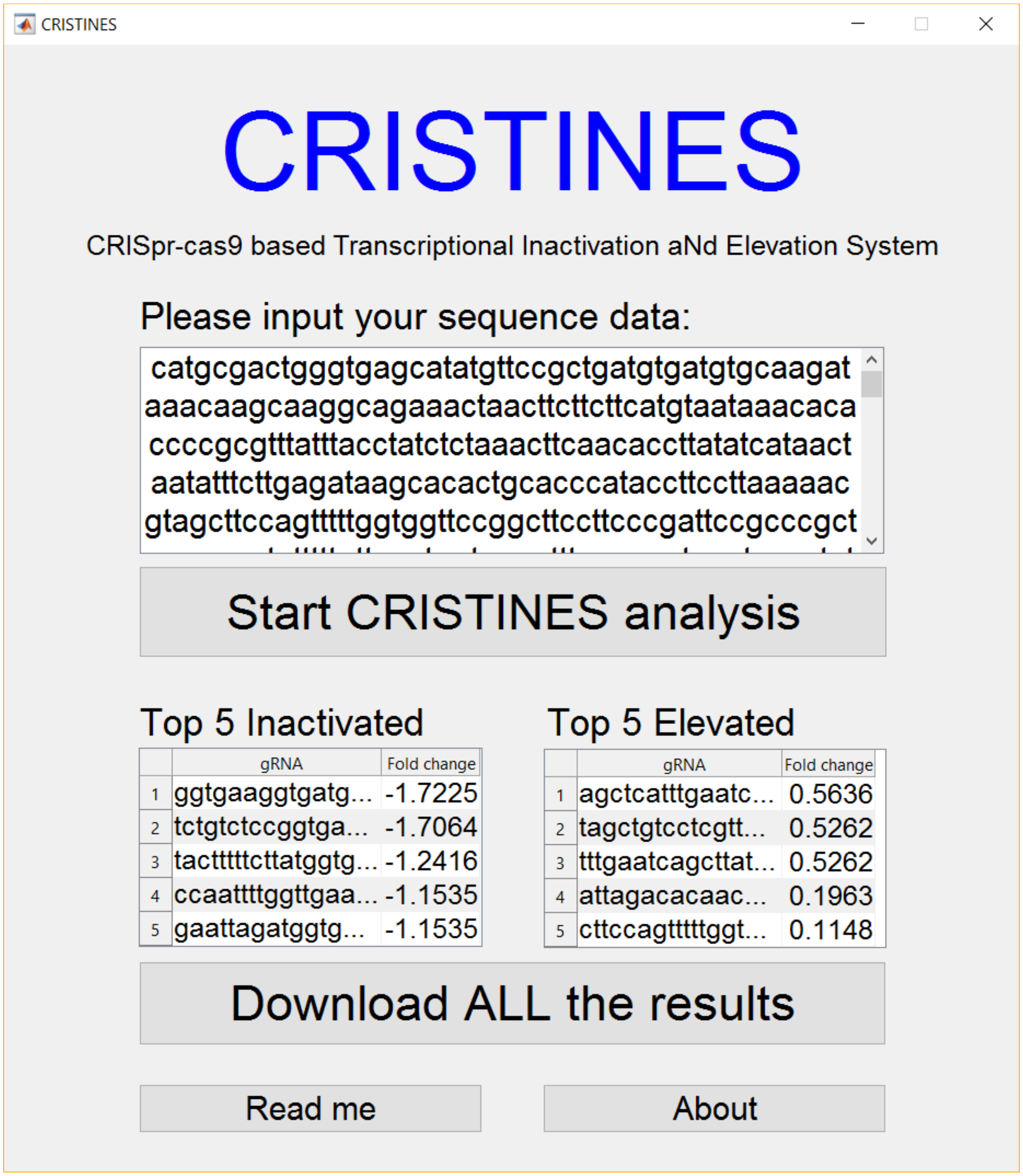

**Figure S3.** Screenshot of CRISTINES software with a demo sequence.

**Table S1.** DNA sequences and plasmids used in this study.

| DNA Name | Sequence (5’-3’) |
|---|---|
| TEF1p | atagcttcaaaatgtttctactcctttttttactcttccagattttctcggactccgcgcatcgccgtaccacttcaaaacacccaagcacagcatactaaatttcccctctttcttcctctagggtgtcgttaattacccgtactaaaggtttggaaaagaaaaaagagaccgcctcgtttctttttcttcgtcgaaaaaggcaataaaaatttttatcacgtttctttttcttgaaaattttttttttgatttttttctctttcgatgacctcccattgatatttaagttaataaacggtcttcaatttctcaagtttcagtttcatttttcttgttctattacaacttttttttacttcttgctcattagaaagaaagcatagcaatctaatctaagttttaattacaaa |
| ENO2p | gtgtcgacgctgcgggtatagaaagggttctttactctatagtacctcctcgctcagcatctgcttcttcccaaagatgaacgcggcgttatgtcactaacgacgtgcaccaacttgcggaaagtggaatcccgttccaaaactggcatccactaattgatacatctacacaccgcacgcctttttttctgaagcccactttcgtggactttgccatatgcaaaattcatgaagtgtgataccaagtcagcatacacctcactagggtagtttctttggttgtattgatcatttggttcatcgtggttcattaatttttttttctccattgctttctggctttgatcttactatcatttggatttttgtcgaaggttgtagaattgtatgtgacaagtggcaccaagcatatataaaaaaaaaaaagcattatcttcctaccagagttgattgttaaaaacgtatttatagcaaacgcaattgtaattaattcttattttgtatcttttcttcccttgtctcaatcttttatttttattttatttttcttttcttagtttctttcataacaccaagcaactaatactataacatacaataata |
| PDC1p | catgcgactgggtgagcatatgttccgctgatgtgatgtgcaagataaacaagcaaggcagaaactaacttcttcttcatgtaataaacacaccccgcgtttatttacctatctctaaacttcaacaccttatatcataactaatatttcttgagataagcacactgcacccataccttccttaaaaacgtagcttccagtttttggtggttccggcttccttcccgattccgcccgctaaacgcatatttttgttgcctggtggcatttgcaaaatgcataacctatgcatttaaaagattatgtatgctcttctgacttttcgtgtgatgaggctcgtggaaaaaatgaataatttatgaatttgagaacaattttgtgttgttacggtattttactatggaataatcaatcaattgaggattttatgcaaatatcgtttgaatatttttccgaccctttgagtacttttcttcataattgcataatattgtccgctgcccctttttctgttagacggtgtcttgatctacttgctatcgttcaacaccaccttattttctaactattttttttttagctcatttgaatcagcttatggtgatggcacatttttgcataaacctagctgtcctcgttgaacataggaaaaaaaaatatataaacaaggctctttcactctccttgcaatcagatttgggtttgttccctttattttcatatttcttgtcatattcctttctcaattattattttctactcataacctcacgcaaaataacacagtcaaatcaatcaaa |
| PGK1p | tcaggcatgaacgcatcacagacaaaatcttcttgacaaacgtcacaattgatccctccccatccgttatcacaatgacaggtgtcattttgtgctcttatgggacgatccttattaccgctttcatccggtgatagaccgccacagaggggcagagagcaatcatcacctgcaaacccttctatacactcacatctaccagtgtacgaattgcattcagaaaactgtttgcattcaaaaataggtagcatacaattaaaacatggcgggcacgtatcattgcccttatcttgtgcagttagacgcgaatttttcgaagaagtaccttcaaagaatggggtctcatcttgttttgcaagtaccactgagcaggataataatagaaatgataatatactatagtagagataacgtcgatgacttcccatactgtaattgcttttagttgtgtatttttagtgtgcaagtttctgtaaatcgattaatttttttttctttcctctttttattaaccttaatttttattttagattcctgacttcaactcaagacgcacagatattataacatctgcacaataggcatttgcaagaattactcgtgagtaaggaaagagtgaggaactatcgcatacctgcatttaaagatgccgatttgggcgcgaatcctttattttggcttcaccctcatactattatcagggccagaaaaaggaagtgtttccctccttcttgaattgatgttaccctcataaagcacgtggcctcttatcgagaaagaaattaccgtcgctcgtgatttgtttgcaaaaagaacaaaactgaaaaaacccagacacgctcgacttcctgtcttcctattgattgcagcttccaatttcgtcacacaacaaggtcctagcgacggctcacaggttttgtaacaagcaatcgaaggttctggaatggcgggaaagggtttagtaccacatgctatgatgcccactgtgatctccagagcaaagttcgttcgatcgtactgttactctctctctttcaaacagaattgtccgaatcgtgtgacaacaacagcctgttctcacacactcttttcttctaaccaaggggggtggtttagtttagtagaacctcgtgaaacttacatttacatatatataaacttgcataaattggtcaatgcaagaaatacatatttggtcttttctaattcgtagtttttcaagttcttagatgctttctttttctctttttttacagatcatcaaggaagtaattatctactttttacaacaaatataaaaca |
| TPI1p | tatatctaggaacccatcaggttggtggaagattacccgttctaagacttttcagcttcctctattgatgttacacctggacaccccttttctggcatccagtttttaatcttcagtggcatgtgagattctccgaaattaattaaagcaatcacacaattctctcggataccacctcggttgaaactgacaggtggtttgttacgcatgctaatgcaaaggagcctatatacctttggctcggctgctgtaacagggaatataaagggcagcataatttaggagtttagtgaacttgcaacatttactattttcccttcttacgtaaatatttttctttttaattctaaatcaatcttttcaatttttttgtttgtattcttttcttgcttaaatctataactacaaaaaacacatacataaactaaaa |
| EGFP | atgtctaaaggtgaagaattattcactggtgttgtcccaattttggttgaattagatggtgatgttaatggtcacaaattttctgtctccggtgaaggtgaaggtgatgctacttacggtaaattgaccttaaaatttatttgtactactggtaaattgccagttccatggccaacctagtcactactttaacttatggtgttcaatgtttttctagatacccagatcatatgaaacaacatgactttttcaagtctgccatgccagaaggttatgttcaagaaagaactattttttttcaaagatgacggtaactacaagaccagagctgaagtcaagtttgaaggtgataccttagttaatagaatcgaattaaaaggtattgattttaaagaagatggtaacattttaggtcacaaattggaatacaactataactctcacaatgtttacatcatggctgacaaacaaaagaatggtatcaaagttaacttcaaaattagacacaacattgaagatggttctgttcaattagctgaccattatcaacaaaatactccaattggtgatggtccagtcttgttaccagacaaccattacttatccactcaatctgccttatccaaagatccaaacgaaaagagagaccacatggtcttgttagaatttgttactgctgctggtattacccatggtatggatgaattgtacaaataa |
| tdTomato | atggtgagcaagggcgaggaggtcatcaaagagttcatgcgcttcaaggtgcgcatggagggctccatgaacggccacgag |
| | ttcgagatcgagggcgagggcgagggccgcccctacgagggcacccagaccgccaagctgaaggtgaccaagggcggc<br>cccctgcccttcgcctgggacatcctgtcccccagttcatgtacggctccaaggcgtacgtgaagcaccccgccgacatccc<br>cgattacaagaagctgtccttccccgagggcttcaagtgggagcgcgtgatgaacttcgaggacggcggtctggtgaccgtga<br>cccaggactcctccctgcaggacggcacgctgatctacaaggtgaagatgcgcggcaccaacttcccccccgacggccccg<br>taatgcagaagaagaccatgggctgggaggcctccaccgagcgcctgtaccccgcgacggcgtgctgaagggcgagatc<br>caccaggccctgaagctgaaggacggcggccactacctggtggagttcaagaccatctacatggccaagaagcccgtgcaa<br>ctgcccggctactactacgtggacaccaagctggacatcacctcccacaacgaggactacaccatcgtggaacagtacgagc<br>gctccgagggccgccaccacctgttcctggggcatggcaccggcagcaccggcagcggcagctccggcaccgcctcctcc<br>gaggacaacaacatggccgtcatcaaagagttcatgcgcttcaaggtgcgcatggagggctccatgaacggccacgagttcg<br>agatcgagggcgagggcgagggccgcccctacgagggcacccagaccgccaagctgaaggtgaccaagggcggccccc<br>tgcccttcgcctgggacatcctgtcccccagttcatgtacggctccaaggcgtacgtgaagcaccccgccgacatccccgatt<br>acaagaagctgtccttccccgagggcttcaagtgggagcgcgtgatgaacttcgaggacggcggtctggtgaccgtgaccca<br>ggactcctccctgcaggacggcacgctgatctacaaggtgaagatgcgcggcaccaacttcccccccgacggccccgtaatg<br>cagaagaagaccatgggctgggaggcctccaccgagcgcctgtaccccgcgacggcgtgctgaagggcgagatccacca<br>ggccctgaagctgaaggacggcggccactacctggtggagttcaagaccatctacatggccaagaagcccgtgcaactgcc<br>cggctactactacgtggacaccaagctggacatcacctcccacaacgaggactacaccatcgtggaacagtacgagcgctcc<br>gagggccgccaccacctgttcctgtacggcatggacgagctgtacaagtaa |
| Sapphire | atgtctaaaggtgaagaattattcactggtgttgtcccaattttggttgaattagatggtgatgttaatggtcacaaattttctgtctcc<br>ggtgaaggtgaaggtgatgctacttacggtaaattgaccttaaaatttatttgtactactggtaaattgccagttccatggccaacct<br>tagtcactactttttcttatggtgttatggtttttgctagatacccagatcatatgaaacaacatgactttttcaagtctgccatgccaga<br>aggttatgttcaagaaagaactatttttttcaaagatgacggtaactacaagaccagagctgaagtcaagtttgaaggtgataccttt<br>agttaatagaatcgaattaaaaggtattgattttaaagaagatggtaacattttaggtcacaaattggaatacaactttaactctcaca<br>atgtttacatcatggctgacaaacaaaagaatggtatcaaagctaacttcaaaattagacacaacattgaagatggtggtgttcaat<br>tagctgaccattatcaacaaaatactccaattggtgatggtccagtcttgttaccagacaaccattacttatccattcaatctgcctta<br>tccaaagatccaaacgaaaagagagaccacatggtcttgttagaatttgttactgctgctggtattacccatggtatggatgaattg<br>tacaaataa |
| mCherry | atggttagcaaaggcgaggaagataacatggctataatcaaagagtttatgagattcaaagtacacatggagggttcagtgaat<br>ggtcatgaatttgaaattgaaggcgaaggcgagggcagaccttacgaaggaactcaaacagcaaaacttaaggtaacaaaag<br>gtggtcctctgccattcgcctgggacattctcagtccacaattcatgtacggttctaaagcgtacgtcaaacatccagcagacatt<br>ccagattacttgaaattgtcttttccagaaggctttaagtgggaaagagttatgaacttcgaggatggagggggttgtgaccgttac<br>gcaagattcctctttacaagatggtgagtttatctacaaggtcaaattaagggggactaattttccttcagacgggccagtcatgca<br>gaaaaagactatgggatgggaagcctcttcagagagaatgtatcctgaagatggcgctctaaaaggagaaatcaagcaaagat<br>tgaagttaaaggacggaggtcattatgatgcagaagtaaaaacaacctataaagctaaaaagccagttcaacttcctggtgccta<br>caatgttaacatcaagctagacattacatcccataatgaagattacactatagtggaacagtatgaacgtgctgaaggtagacac<br>agtacaggtggtatggatgaactgtacaagtaa |
| Venus | atgtctaaaggtgaagaattattcactggtgttgtcccaattttggttgaattagatggtgatgttaatggtcacaaattttctgtctcc<br>ggtgaaggtgaaggtgatgctacttacggtaaattgaccttaaaattgatttgtactactggtaaattgccagttccatggccaacc<br>ttagtcactactttaggttatggtttgcaatgttttgctagatacccagatcatatgaaacaacatgactttttcaagtctgccatgcca<br>gaaggttatgttcaagaaagaactatttttttcaaagatgacggtaactacaagaccagagctgaagtcaagtttgaaggtgatac<br>cttagttaatagaatcgaattaaaaggtattgattttaaagaagatggtaacattttaggtcacaaattggaatacaactataactctc<br>acaatgtttacatcactgctgacaaacaaaagaatggtatcaaagctaacttcaaaattagacacaacattgaagatggtggtgtt<br>caattagctgaccattatcaacaaaatactccaattggtgatggtccagtcttgttaccagacaaccattacttatcctatcaatctgc<br>cttatccaaagatccaaacgaaaagagagaccacatggtcttgttagaatttgttactgctgctggtattacccatggtatggatga<br>attgtacaaataa |
| VioA | atgaagcacagttctgatatttgcatagttggggcgggcatctcagggttgacctgcgcgagtcaccttcttgactctcctgcttgt<br>cgtggcttatccctaagaatttttgatatgcaacaggaggctggaggtaggatcaggtctaagatgcttgatggcaaagcgagca<br>ttgagttgggcgctggccgttattcccctcaacttcatcctcacttccaaagcgcgatgcaacactattcacagaagagtgaggttt<br>atccgtttacacaattaaaattcaagagccatgtccagcagaagctaaagagagctatgaatgagctttccccccgtctaaagga<br>gcacggaaaagaatcctttctacagtttgtgtctcgttaccaaggtcacgactccgcggtcggcatgatcaggtctatgggatatg<br>acgcgctattccttccggacatatctgcagaaatggcatacgatattgttggcaaacatcccgagatccaaagtgttactgacaat<br>gacgctaatcagtggtttgccgcagaaactgggttcgccgggcttattcaggggattaaggcgaaagtaaaggctgcaggtgc<br>tagattttccctggggtataggttactgtcagtgaggacggacggggacggatacttgctgcaattagctggtgatgatggctgg<br>aaacttgagcatagaacaagacatctaattcttgcaatacctccatcagctatggcgggactaaatgtagattttcctgaagcgtg<br>gtcaggggccaggtatgggtcattaccgctatttaaaggcttcttaacttacggagagccatggtggttagattataagctagatg |
| | accaggtgctaattgtggataatccgcttagaaagatctactttaaaggcgataagtatttattttttacacagactcagaaatggc<br>aaattattggagaggctgtgtagcggagggcgaggatggttatttggaacaaatccgtactcacttggcctcagcgctaggaat<br>cgttagggagcgtataccccaaccgttagcccatgttcacaagtattgggcacatggtgtcgagttttgcagggattcagatattg<br>accacccatcagcccttagccaccgtgactccggaatcattgcgtgcagtgacgcctatactgaacattgcggatggatggagg<br>ggggtcttttatctgcgagagaagccagcaggctgcttctacaaaggatagccgcctaa |
| VioB | atgagcattctggattttccacgcatccatttccgcggctgggcgcgggtcaacgcgcccaccgccaaccgcgatccgcacgg<br>ccacatcgacatggccagcaatacggtggccatggcaggcgaaccgttcgacctcgcgcgccatccgaccgagttccaccg<br>ccacctgcggtcgctggggccgcgtttcggcctggacggccgggctgacccggaagggccgttcagcctggccgagggct<br>acaacgcggccggcaacaaccatttctcctgggagagcgccaccgtcagccacgtgcagtgggatggcggcgaagcggac<br>cgcggcgacggcctggtcggcgccaggctggcgctgtgggggcattacaacgattacctgcgcaccaccttcaaccgcgcg<br>cgctgggtggacagcgaccccacccgccgcgacgcggcgcagatctacgccgggcagttcacgatcagcccggccggcg<br>ccggaccgggcacgccctggctgttcaccgccgacatcgacgacagccacggcgcgcgctggacgcgcggcggccacat<br>cgccgagcgcggcggccatttcctggacgaggagttcggcctggcgcggctgttccagttctcggtgcccaaagaccatccg<br>cacttcctgttccacccggggccattcgattccgaagcctggcgcaggctgcagctggcgctggaggacgacgacgtgctcg<br>gcctgacggtgcagtacgcgctgttcaatatgtcgacgccgccgcaacccaactcgccggtgttccacgacatggtcggcgtg<br>gtcggcctgtggcggcgcggcgaactggccagctacccggccggccggctgctgcgtccgcgccagcccgggctgggcg<br>atctgacgctgcgcgtaagcggcggccgcgtggcgctgaatctggcctgcgccattccgttctccacccgggcggcgcagcc<br>gtccgcgccggacaggctgacgcccgatctcggggccaagctgccgttgggcgacctgctgctgcgcgacgaggacggcg<br>cgttgctggcgcgggtgccgcaggcgctttaccaggattactggacgaaccacggcatcgtcgacctgccgctgctgcgcga<br>gcccaggggctcgctgacgctgtccagcgagctggccgaatggcgcgagcaggactgggtcacgcagtccgacgcctcca<br>atctttatttggaagcgccggaccgccgccacggccgtttctttccggaaagcatcgcgctgcgcagctatttccgcggcgagg<br>cccgcgcgcgcccggacattccccaccggatcgaggggatgggtctggtcggcgtggagtcgcgccaggacggcgatgcc<br>gccgaatggcggctgaccggcctgcggcccggcccggcgcgcatcgtgctcgacgacggcgcggaggcgatcccgctgc<br>gggtgctgccggacgactgggcgttggacgacgcgacggtggaggaggtcgattacgccttcctgtaccggcacgtgatgg<br>cctattacgagctggtctacccgttcatgtccgacaaggtgttcagcctggccgaccgctgcaagtgcgagacctacgccagg<br>ctgatgtggcagatgtgcgatccgcagaaccggaacaagagctactacatgcccagcacccgcgagctgtcggcgcccaag<br>gccaggctgttcctcaaatacctggcccatgtcgagggccaggccaggctgcaggcgccgccgccggccgggccggcgcg<br>catcgagagcaaggcccagctggcggccgagctgcgcaaggcggtggatctggagttgtcggtgatgctgcagtacctgtac<br>gccgcctattccattcccaattacgcccagggccagcagcgggtgcgcgacggcgcgtggacggcggagcagctgcagct<br>ggcctgcggcagcggcgaccggcgccgcgacggcggcatccgcgccgcgctgctggagatcgcccacgaggagatgatc<br>cattacctggtggtcaacaacctgctgatggcgctgggcgagccgttctacgccggcgtgccgctgatgggcgaggcggcgc<br>ggcaggcgttcggcctggacaccgaattcgcgctggagccgttctccgagtcgacgctggcgcgcttcgtccggctggaatg<br>gccgcacttcatccctgcgccgggcaaatccatcgccgactgctacgccgccatccgccaggcctttctcgatctgcccgacct<br>gttcggcggcgaggccggcaagcgcggcggcgagcaccacttgttcctcaacgagctgaccaaccgcgcccatcccggct<br>accagctggaggtgttcgatcgcgacagcgcgctgttcggcatcgccttcgtcaccgaccagggcgagggcggggcgctgg<br>actcgccgcattacgagcattcgcatttccagcggctgcgggagatgtcggccaggatcatggcgcagtccgcgccgttcgag<br>ccggcgttgccggcgctgcgcaacccggtgctggacgagtcgccgggctgccagcgcgtggcggacggacgggcgcgc<br>gcgctgatggcgctgtaccagggcgtgtacgagctgatgttcgcgatgatggcgcagcacttcgcggtcaagccgctgggca<br>gcctcaggcgctcgcggctgatgaacgcggcgatcgacctgatgaccggcctgctcaggccgctgtcctgcgcgctgatgaa<br>cctgccgtcgggcatcgccggacgcaccgccgggccgccgctgccggggccggtggatacccgcagctacgacgactac<br>gcgctgggctgccggatgctggcgcggcgctgcgagcgcctgctggagcaggcgtcgatgctggagccgggctggctgcc<br>cgacgcgcaaatggaactgctggatttctaccgccggcagatgctggatttggcttgtggaaagctttctagagaggcctaa |
| VioC | atgcagaaagccattatcgtgggaggcggcttagccggtggcttaacagctatttacctagcaaggcgtggctatgaggtacac<br>gtagtagagaagaggggtgaccccttgcaagatctgtcatcttatgtcgacgcagtttccagtcgtgctataggggtaagtatga<br>cggtcaggggcattaaagcggttttggccgctggtatccccagggctgagctggatcaatgtggagagccgatagttgggatg<br>gcatttagtgtgggaggaaagttcaggatacgtgaattacaacctctggaagggctttctcccttgtccctgaatagggcagcctt<br>tcagagactgttaaatcgtcacgcgaatcgtaatggggtgaggtatcatttcgagcacaaatgtcttgacgtcgacttggaaggc<br>aagtccgtactgattcaaagcaaggacgggcaactacaaagactgcaaggtgatatggtaattggcgccgacggcgcacata<br>gcgccgtaaggcaggcaatgcaaagtggtatgcgtaggttcgagttccaacagactttcttccgtcatggttataaaaccttagtt<br>ctgccggatgccgaggctctgggttatagaaaagatttgttgtacttcttcggtatggattctggcggtctattcgcagggagagc<br>agctaccatcccggacgggtcagtcagcatcgccgtgtgtctgccttatgaagggacgccctcccttgctacgcaggatagac<br>aggcgatgaaagctttcttccaaaggtattttggttctctatcagacgcagttagggaggagatgttagagcaattcctagtgaaa<br>cccagcaacgacctaataaatgttaggtccagcaccttccactacaaagggaacgttttattgcttggcgatgccgcccatgcca<br>ctgcgccatttttgggccagggaatgaacatggcgctggaagacgcccacagcttcgtgaccctacttgatcaacattgtcacg<br>atcaggagctggcattcgctgagtttacagcccaacgtaaagtgcaagctgatgcaatgcaagacatggcgcgtgcaaattac |
| | gatgtgttatcctgttcaaatcccatcttctttctgagggcgagatatactaggtacatgcacggaaaattcccggagctttatccac<br>cagatatggcgcaaaagctatattttacgagtgaaccttacgacaaactacaaacaatacaaaagaaacagaacgtttggtacaa<br>gctaggcagggtaaac |
| VioD | atgaagaagattttagttattggggctggacccgcgggtctggtgttcgcatcacagatgaagctagcgaaaccagattggcaa<br>attagcatagctgagaagaacgatccggaggaggtcgctggttggggtgtagttttgccggggcgtcctggccagcatcctgc<br>caaccctttgagctatttggagcatccggagaggcttgatccgcagttcctggaagacttcaaacttatacatcataacgaaccta<br>acctgatgtctactggagtcctactttgcggggttgaacgtagggggcttgtccaggctctgcgtgagaggtgccagtctttggg<br>gatcgcgatccactacgagagcccattactagccagggaacaattacctcttgatgactatgacttaatagtagctgcaaacggg<br>gtgaatcataaaactagccacttcacggaagcgttggcaccgaaactggaatacgggggcaacaaatatatatggttcggtacc<br>tcacaactgtttgaccaaatgaacttggtatttaggacacacggaaaagatatattcatcgcgcacgcatataagtattcaagcag<br>aatgagtacattcgttgtcgagtgttccgaagaaaccttcgagcgtgcaaggttaggtgagatgtctgacgaagctagtgccga<br>gtacgtagccggggtcttcagggcggagcttggtggtcacgggctggtcgcacaacctgggcttgggtggcgtaatttcatga<br>ccctttcacacgataagtcttacgacggaaagttagttttaatcggcgacgctctgcaaagcggacacttcagtattgggcatggc<br>acaacgatggcagtggtggctgcccagcttctagtcaaagctctgtgcgctgaagccagcgtccctgccgctctagcctcatttg<br>acgcgagagcgatgccattagtaaagctgttttccgagcacgcgaattccagccgtgattggttcgagacggtcgatgatagaa<br>tgcacctgtcaaacgcggaattcatgcagtccttcgacgcaagaaggaaagctttgcctcccttacccgaggcgttggcccgta<br>atctaggctacgccttagataga |
| VioE | atggaaaatagagagcccccactgttaccagctcgttggtctagcgcatacgtgtcctattggtccccgatgctacccgacgacc<br>agctgacatctggttattgctggtttgactatgaaagagatatatgcagaatcgatgggctatttaatccatggagcgaacgtgac<br>accgggtatcgtttatggatgagtgaggtaggtaatgccgcgtctggcaggacctggaaacaaaaggtagcatacggtaggga<br>gagaactgctcttggggagcagctgtgtgaacgtccactggacgatgaaactggcccgtttgctgaactttttttgccgagagat<br>gttctgcgtagactaggcgcacgtcacataggacgtagagttgtattaggcagagaagccgatggatggaggtatcagaggcc<br>gggaaagggcccatccaccctatatttggacgcagcatcaggcacccccttgagaatggtgacgggagacgaggccagtcgt<br>gcgagtctaagagactttcccaatgtgagtgaagcagagatacctgatgctgtgttcgctgccaagcgctaa |
| dCas9_VP64 | atgtctagagccccaaagaagaagagaaaagttagacccggggataagaaatactctattggtttggctatcggtacaaactct<br>gttggttgggctgttattactgatgaatacaaggttccatccaagaagttcaaggttttgggtaacactgatagacactccatcaaa<br>aagaacttgattggtgccttgttgttcgattctggtgaaactgctgaagctactagattgaaaagaaccgctagaagaagatacac<br>cagaagaaagaacagaatctgctacttgcaagaaatcttctccaacgaaatggccaaggttgatgattcattcttccacagattgg<br>aagaatccttcttggtcgaagaagataagaagcacgaaagacatccaatcttcggtaacatcgttgatgaagttgcttaccacga<br>aaagtacccaactatctaccatttgagaaagaagttggttgactctaccgataaggctgatttgagattgatctatttggctttggcc<br>cacatgattaagttcagaggtcatttcttgatcgaaggtgatttgaacccagataactccgatgttgataagttgttcatccaattagt<br>ccaaacctacaatcaattattcgaagaaaacccaatcaacgcctctggtgttgatgctaaagctattttgtctgccagattgtccaa<br>gtccagaagattagaaaatttgatcgcccaattaccaggtgaaaagaagaatggtttgttcggtaatttgattgccttgtctttgggt<br>ttgactccaaacttcaagtccaatttcgatttggctgaagatgccaagttgcaattatctaaggatacctacgatgacgatttggata<br>acttgttggctcaaatcggtgatcaatacgctgatttgtttttggctgctaagaacttgtccgatgccattttgttgtccgatattttgag<br>agtcaacaccgaaattactaaggctccattgtctgcctctatgatcaaaagatacgatgaacaccaccaagacttgactttgttga<br>aggctttggtcagacaacaattacctgaaaagtacaaagaaattttcttcgatcaatccaagaacggttacgccggttatattgatg<br>gtggtgcttctcaagaagaattttacaagttcatcaagccaatcttggaaaagatggacggtactgaagaattattggtcaagttga<br>acagagaagatttgttgagaaagcaaagaaccttcgacaacggttctattccacatcaaattcacttgggtgaattgcacgcaatt<br>ttgagaagacaagaagatttttatccattcttgaaggacaacagagaaaagatcgaaaagattctgaccttcagaatcccttacta<br>cgttggtccattggctagaggtaattcaagatttgcctggatgactagaaagtccgaagaaactattactccttggaacttcgaag<br>aagttgtagataagggtgcttctgcccaatcctttattgaaagaatgaccaacttcgacaagaacttgccaaacgaaaaggttttg<br>ccaaagcactctttgttgtacgaatacttcaccgtctacaacgaattgactaaggttaagtacgtcaccgaaggtatgagaaaacc<br>agctttttatccggtgaacaaaagaaggctatcgtcgatttgttgttcaagaccaacagaaaggttactgtcaagcaattaaaaga<br>agattacttcaagaaaatcgaatgcttcgactccgttgaaatttctggtgtcgaagatagattcaatgcctctttaggtacttaccatg<br>acttgttgaaaatcatcaaggacaaggatttcttggacaacgaagaaaacgaagatattttggaagatattgtcttgacattgacct<br>tgtttgaagatagagaaatgattgaagaaagattgaaaacctacgcccacttgttcgatgataaggttatgaagcaattaaagaga<br>agaagatacactggttggggtagattgtccagaaaattgattaacggtatcagagacaagcaatccggtaagaccattttggact<br>tttgaagtctgatggtttcgctaacagaaacttcatgcaattaatccacgacgattccttgactttcaaagaagatatacaaaaggc<br>ccaagtctctggtcaaggtgattctttacatgaacatatcgctaacttggctggttctccagctattaagaagggtattttacaaacc<br>gttaaggtcgttgacgaattggtcaaagttatgggtagacataagccagaaaacatcgttatcgaaatggctagagaaaatcaaa<br>ccacccaaaagggtcaaaagaactccagagaaagaatgaagagaatcgaagaaggtatcaaagaattgggttcccaaattttg<br>aaagaacacccagttgaaaacacccaattacaaaacgaaaagttgtacttgtactacttgcaaaacggtagagatatgtacgttg<br>accaagaattggacatcaacagattgtctgattacgatgttgacgctatcgttccacaatctttttttgaaggatgactccattgacaa<br>caaggtcttgactagatccgataagaatagaggtaagtccgataacgttccatctgaagaagtcgttaagaaaatgaagaactat |
| | tggagacaattattgaacgccaagttgatcacccaaagaaagtttgacaatttgaccaaggctgaaagaggtggtttgtctgaattggataaggcaggtttcatcaaaagacaattagtagaaaccagacaaatcaccaagcacgttgctcaaattttggatagtagaatgaacactaagtacgacgaaaacgacaaattgatcagagaagttaaggtcattaccttgaagtccaagttggtttccgatttcagaaaggacttccaattctacaaggtcagagaaatcaacaactaccatcatgcacatgatgcttacttgaatgctgttgttggtactgccttgattaagaagtatccaaagttggaatccgaatttgtctacggtgattacaaggtttacgacgttagaaagatgatcgccaagtccgaacaagaaattggtaaagctactgccaaatacttcttctactccaatattatgaatttctttaagaccgaaatcactttggccaacggtgaaattagaaaaagaccattgattgaaactaatggtgaaacaggtgaaatcgtttgggataagggtagagattttgccactgttagaaaggtattgtccatgccacaagtaaacatcgtcaaaaagaccgaagttcaaactggtggtttctccaaagaatccattttgcctaagagaaactccgataagttgatcgctagaaaaaaagactgggacccaaaaaagtacggtggttttgattctccaactgttgcttactctgttttggttgttgctaaggtcgaaaagggtaagagtaagaagttgaagtccgtcaaagaattattaggtatcactatcatggaaagatcctcattcgaaaagaatcctatcgactttttggaagccaagggttacaaagaagtcaagaaggacttgatcattaagttgccaaagtacagtttgttcgaattggaaaatggtagaaagagaatgttggcttctgccggtgaattacaaaagggtaatgaattggcttgccatccaagtacgttaatttcttatacttggcctcccactacgaaaaattgaaaggttctcctgaagataacgaacaaaagcaattatttgtcgaacaacacaagcactacttggacgaaatcattgaacaaatttccgaattttccaaaagagtcattttggctgacgccaatttggacaaagttttgtcagcttacaacaagcacagagataagccaattagagaacaagctgaaaacatcattcacttgttcacttgactaacttgggtgctccagctgcttttaagtatttcgataccactatcgacagaaagagatacacctctaccaaagaagttttggacgctactttgatccaccaatctattactggtttgtacgaaactagaatcgacttgtctcaattaggtggtgatggttctggtagatctggagtcgacggtggaggttctgacgctttggacgacttcgacttggatatgctgggttctgatgcgctagatgactttgacctcgacatgcttggaagtgacgccttagatgattttgacctggatatgcttggatcagacgctctggacgatttcgacttagacatgctttcctaa |
| gRNA expression cassette | gagctcggggatctgccaattgaacataacatggtagttacatatactagtaatatggttcggcacacattaaaagtataaaaactatctgaattacgaattacatatattggtcataaaaatcaatcaatcatcgtgtgttttatatgtctcttatctaagtataagaatatccatagttaatattcacttacgctaccttttaacctgtaatcattgtcaacaggatatgttaacgacccacattgataaacgctagtatttcttttcctcttcttattggccggctgtctctatactcccctatagtctgtttcttttcgtttcgattgttttacgtttgaggcctcgtggcgcacatggtacgctgtggtgctcgcggctgggaacgaaactctgggagctgcgattggcagaagcttNNNNNNNNNNNNNNNNNNNNNgttttagagctagaaatagcaagttaaaataaggctagtccgttatcaacttgaaaaagtggcaccgagtcggtgctttttttctcgagccatatccaacttccaatttaatctttctttttaattttcacttatttgcgatacagaaagaaaaaagcgatagtaactattgaattttgtttggatttggttagattagatatggtttctctttatatttacatgctaaaaatgggctacaccagagatacataattagatatatatacgccagtacaccttatcggcccaagccttgtcccaaggcagcgttttgttcttggaaacgctgccctacacgttcgctatgcttcaagaacttttctgagcacttcatgatgcatgtttgttccttattggttagctttgatgttgtgaagtcattgacacagtctgtgaaacatctttctaccagattagagtacaaacgcatgaaatccttcatttgcttttgttccactactttttggaactcttgttgttctttgg |
| **Plasmid Name** | **Description** |
| p4FP | pRS416-tef1p-GFP-tef1t-pdc1p-sapphire-pdc1t-tpi1p-mCHerry-tpi1t-pgk1p-Venus-pgk1t |
| pTomato | pRS416-eno2p-tdTomato-eno2t |
| pViolacein | pRS416- tef1p-VioA-tef1t- pgk1p-VioB-pgk1t- eno2p-VioC-eno2t- tpi1p-VioD-cyc1t-pdc1p-VioE-pdc1t |
| pCas9 | pRS413-gal1p-dCas9_VP64-cyc1t |
| pgRNA(n)* | pRS425-rpr1p-gRNA-rpr1t |

* All the gRNAs use the same structure except the 20bp targeting to different sites.

**Table S2.** Strains used in this study.

| Name | Phenotype | Plasmids |
|---|---|---|
| TEF1p(H) | INVSc1: MATa his3D1 leu2 trp1-289 ura3-52 MAT his3D1 leu2 trp1-289 ura3-52 | pViolacein, pCas9, pgRNA-TEF1p(H) |
| TEF1p(M) | Same as INVSc1 | pViolacein, pCas9, pgRNATEF1p(M) |
| TEF1p(L) | Same as INVSc1 | pViolacein, pCas9, pgRNATEF1p(L) |
| PDC1p(H) | Same as INVSc1 | pViolacein, pCas9, pgRNAPDC1p(H) |
| PDC1p(M) | Same as INVSc1 | pViolacein, pCas9, pgRNAPDC1p(M) |
| PDC1p(L) | Same as INVSc1 | pViolacein, pCas9, pgRNAPDC1p(L) |
| TPI1p(M) | Same as INVSc1 | pViolacein, pCas9, pgRNATPI1p(M) |
| TPI1p(L) | Same as INVSc1 | pViolacein, pCas9, pgRNATPI1p (L) |
| ENO2p(H) | Same as INVSc1 | pViolacein, pCas9, pgRNAENO2p(H) |
| ENO2p(M) | Same as INVSc1 | pViolacein, pCas9, pgRNAENO2p(M) |
| ENO2p(L) | Same as INVSc1 | pViolacein, pCas9, pgRNAENO2p(L) |
| ACD | Same as INVSc1 | pViolacein, pCas9, pgRNATEF1p(H)-gRNAENO2p(H)-gRNATPI1p(L) |
| CD | Same as INVSc1 | pViolacein, pCas9, pgRNAENO2p(H)-gRNATPI1p(L) |
| AD | Same as INVSc1 | pViolacein, pCas9, pgRNATEF1p(H)-gRNATPI1p(L) |

**Table S3.** Key parameters of guide RNAs used in this study for model training.

TEF1p-GFP

ATAGCTTCAAAATGTTTCTACTCCTTTTTTACTCTTCCAGATTTTCTCGGACTCCGCGC

ATCGCCGTACCACTTCAAAACACCCAAGCACAGCATACTAAATTTCCCCTCTTTCTTC

CTCTAGGGTGTCGTTAATTACCCGTACTAAAGGTTTGGAAAAGAAAAAAGAGACCGC

CTCGTTTCTTTTTCTTCGTCGAAAAAGGCAATAAAAATTTTTATCACGTTTCTTTTTCT
G1
TGAAAATTTTTTTTTTTGATTTTTTTCTCTTTCGATGACCTCCCATTGATATTTAAGTTA

ATAAACGGTCTTCAATTTCTCAAGTTTCAGTTTCATTTTTCTTGTTCTATTACAACTTTT
G2
TTTACTTCTTGCTCATTAGAAAGAAAGCATAGCAATCTAATCTAAGTTTTAATTACAAA

**ATG**TCTAAAGGTGAAGAATTATTCACTGGTGTTGTCCCAATTTTGGTTGAATTAGATG
G3 G4 G5 G6
GTGATGTTAATGGTCACAAATTTTCTGTCTCCGGTGAAGGTGAAGGTGATGCTACTTA
G7 G8 G9 G10
CGGTAAATTGACCTTAAAATTTATTTGTACTACTGGTAAATTGCCAGTTCCATGGCCAA
G11 G12 G13
CCTTAGTCACTACTTTAACTTATGGTGTTCAATGTTTTTCTAGATACCCAGATCATATGA
G14
AACAACATGACTTTTTCAAGTCTGCCATGCCAGAAGGTTATGTTCAAGAAAGAACTAT
G15
TTTTTTCAAAGATGACGGTAACTACAAGACCAGAGCTGAAGTCAAGTTTGAAGGTGA
G16 G17
TACCTTAGTTAATAGAATCGAATTAAAAGGTATTGATTTTAAAGAAGATGGTAACATTT
G18 G19
TAGGTCACAAATTGGAATACAACTATAACTCTCACAATGTTTACATCATGGCTGACAA
G20 G21 G22
ACAAAAGAATGGTATCAAAGTTAACTTCAAAATTAGACACAACATTGAAGATGGTTCT
G23 G24
GTTCAATTAGCTGACCATTATCAACAAAATACTCCAATTGGTGATGGTCCAGTCTTGTT
G25
ACCAGACAACCATTACTTATCCACTCAATCTGCCTTATCCAAAGATCCAAACGAAAAG

AGAGACCACATGGTCTTGTTAGAATTTGTTACTGCTGCTGGTATTACCCATGGTATGGA
G26 G27 G28
TGAATTGTACAAA

| No. | Location | GC% | ΔG | AGG | TGG | CGG | GGG | #G | #A | Fold change (log2) |
|---|---|---|---|---|---|---|---|---|---|---|
| G1 | -213 | 0.3 | -9.1 | 1 | 0 | 0 | 0 | 2 | 4 | 0.00 |
| G2 | -114 | 0.1 | -7.3 | 0 | 0 | 1 | 0 | 2 | 9 | -0.02 |
| G3 | 9 | 0.15 | -9.3 | 1 | 0 | 0 | 0 | 1 | 9 | -3.38 |
| G4 | 27 | 0.3 | -12 | 0 | 1 | 0 | 0 | 4 | 9 | -0.29 |
| G5 | 44 | 0.45 | -11 | 0 | 1 | 0 | 0 | 4 | 3 | 0.01 |
| G6 | 57 | 0.3 | -7.8 | 0 | 1 | 0 | 0 | 4 | 6 | 0.07 |
| G7 | 69 | 0.3 | -7 | 0 | 1 | 0 | 0 | 6 | 7 | 0.03 |
| G8 | 90 | 0.4 | -8.4 | 0 | 0 | 1 | 0 | 3 | 4 | -0.24 |
| G9 | 96 | 0.4 | -9.5 | 1 | 0 | 0 | 0 | 4 | 4 | -0.50 |
| G10 | 102 | 0.55 | -10.8 | 1 | 0 | 0 | 0 | 7 | 3 | -1.54 |
| G11 | 117 | 0.45 | -8.8 | 0 | 0 | 1 | 0 | 7 | 5 | -0.09 |
| G12 | 150 | 0.15 | -10.4 | 0 | 1 | 0 | 0 | 1 | 7 | -0.15 |
| G13 | 169 | 0.45 | -12.9 | 0 | 1 | 0 | 0 | 4 | 5 | -0.24 |
| G14 | 198 | 0.25 | -8.6 | 0 | 1 | 0 | 0 | 1 | 6 | 0.07 |
| G15 | 270 | 0.5 | -9.9 | 1 | 0 | 0 | 0 | 4 | 5 | -0.01 |
| G16 | 309 | 0.2 | -8 | 0 | 0 | 1 | 0 | 2 | 7 | -0.19 |
| G17 | 345 | 0.4 | -10 | 1 | 0 | 0 | 0 | 6 | 7 | -0.35 |
| G18 | 378 | 0.2 | -7.1 | 1 | 0 | 0 | 0 | 3 | 10 | -0.04 |
| G19 | 399 | 0.25 | -7.9 | 0 | 1 | 0 | 0 | 5 | 8 | 0.05 |
| G20 | 411 | 0.25 | -10.8 | 1 | 0 | 0 | 0 | 4 | 9 | 0.08 |
| G21 | 422 | 0.25 | -8.1 | 0 | 1 | 0 | 0 | 2 | 8 | -0.97 |
| G22 | 458 | 0.35 | -7.1 | 0 | 1 | 0 | 0 | 1 | 6 | 0.08 |
| G23 | 477 | 0.35 | -7.6 | 0 | 1 | 0 | 0 | 4 | 11 | -0.31 |
| G24 | 519 | 0.3 | -7 | 0 | 1 | 0 | 0 | 3 | 10 | 0.13 |
| G25 | 564 | 0.25 | -8.1 | 0 | 1 | 0 | 0 | 0 | 10 | -0.21 |
| G26 | 653 | 0.4 | -7 | 0 | 1 | 0 | 0 | 4 | 12 | -0.12 |
| G27 | 681 | 0.35 | -8.9 | 0 | 1 | 0 | 0 | 4 | 4 | 0.00 |
| G28 | 693 | 0.5 | -14.5 | 0 | 1 | 0 | 0 | 4 | 4 | 0.21 |

TPI1p-mCherry

TATATCTAGGAACCCATCAGGTTGGTGGAAGATTACCCGTTCTAAGACTTTTCAGCTTC
M1 M2
CTCTATTGATGTTACACCTGGACACCCCTTTTCTGGCATCCAGTTTTTAATCTTCAGTG
M3 M4
GCATGTGAGATTCTCCGAAATTAATTAAAGCAATCACACAATTCTCTCGGATACCACCT
M5
CGGTTGAAACTGACAGGTGGTTTGTTACGCATGCTAATGCAAAGGAGCCTATATACCT
M6 M7 M8 M9
TTGGCTCGGCTGCTGTAACAGGGAATATAAAGGGCAGCATAATTTAGGAGTTTAGTGA
M10 M11 M12 M13 M14
ACTTGCAACATTTACTATTTTCCCTTCTTACGTAAATATTTTTCTTTTTAATTCTAAATCA

ATCTTTTTCAATTTTTTGTTTGTATTCTTTTCTTGCTTAAATCTATAACTACAAAAAACA

CATACATAAACTAAAA**ATGG**TTAGCAAAGGCGAGGAAGATAACATGGCTATAATCAA
M15 M16
AGAGTTTATGAGATTCAAAGTACACATGGAGGGTTCAGTGAATGGTCATGAATTTGAA
M17 M18 M19
ATTGAAGGCGAAGGCGAGGGCAGACCTTACGAAGGAACTCAAACAGCAAAACTTAA
M20

GGTAACAAAAGGTGGTCCTCTGCCATTCGCCTGGGACATTCTCAGTCCACAATTCATG
M21 M22 M23

TACGGTTCTAAAGCGTACGTCAAACATCCAGCAGACATTCCAGATTACTTGAAATTGT
M24

CTTTTCCAGAAGGCTTTAAGTGGGAAAGAGTTATGAACTTCGAGGATGGAGGGGTTG
M25 M26 M27

TGACCGTTACGCAAGATTCCTCTTTACAAGATGGTGAGTTTATCTACAAGGTCAAATT
M28

AAGGGGGACTAATTTTCCTTCAGACGGGCCAGTCATGCAGAAAAAGACTATGGGATG
M29M30 M31 M32

GGAAGCCTCTTCAGAGAGAATGTATCCTGAAGATGGCGCTCTAAAAGGAGAAATCAA
M33

GCAAAGATTGAAGTTAAAGGACGGAGGTCATTATGATGCAGAAGTAAAAACAACCTA
M34 M35

TAAAGCTAAAAAGCCAGTTCAACTTCCTGGTGCCTACAATGTTAACATCAAGCTAGAC

ATTACATCCCATAATGAAGATTACACTATAGTGGAACAGTATGAACGTGCTGAAGGTA
M36

GACACAGTACAGGTGGTATGGATGAACTGTACAAG
M37 M38

| No. | Location | GC% | Energy | AGG | TGG | CGG | GGG | #G | #A | Fold change (log2) |
|---|---|---|---|---|---|---|---|---|---|---|
| M1 | -408 | 0.45 | -11.8 | 0 | 1 | 0 | 0 | 4 | 6 | 2.36 |
| M2 | -405 | 0.55 | -14.5 | 0 | 1 | 0 | 0 | 6 | 5 | 2.31 |
| M3 | -353 | 0.4 | -8.2 | 0 | 1 | 0 | 0 | 2 | 4 | 2.65 |
| M4 | -314 | 0.3 | -7.4 | 0 | 1 | 0 | 0 | 2 | 5 | 2.60 |
| M5 | -265 | 0.35 | -8.2 | 0 | 0 | 1 | 0 | 1 | 8 | 1.79 |
| M6 | -253 | 0.45 | -8.6 | 0 | 0 | 1 | 0 | 2 | 5 | 1.15 |
| M7 | -239 | 0.55 | -10.7 | 1 | 0 | 0 | 0 | 4 | 5 | 0.35 |
| M8 | -236 | 0.55 | -10.5 | 0 | 1 | 0 | 0 | 6 | 5 | 0.76 |
| M9 | -211 | 0.4 | -11.4 | 1 | 0 | 0 | 0 | 4 | 6 | 0.81 |
| M10 | -194 | 0.4 | -14.2 | 0 | 1 | 0 | 0 | 3 | 7 | 1.93 |
| M11 | -189 | 0.45 | -12.2 | 0 | 0 | 1 | 0 | 4 | 4 | 0.10 |
| M12 | -176 | 0.55 | -11.1 | 1 | 0 | 0 | 0 | 6 | 2 | 0.20 |
| M13 | -165 | 0.35 | -7.4 | 1 | 0 | 0 | 0 | 5 | 8 | 0.26 |
| M14 | -150 | 0.3 | -9.6 | 1 | 0 | 0 | 0 | 4 | 8 | 0.92 |
| M15 | 2 | 0.2 | -7 | 0 | 1 | 0 | 0 | 0 | 13 | 0.05 |
| M16 | 29 | 0.45 | -7 | 0 | 1 | 0 | 0 | 6 | 10 | 0.22 |
| M17 | 68 | 0.3 | -7 | 0 | 1 | 0 | 0 | 3 | 9 | -0.29 |
| M18 | 71 | 0.4 | -7.5 | 1 | 0 | 0 | 0 | 5 | 8 | 0.15 |
| M19 | 84 | 0.5 | -7.7 | 0 | 1 | 0 | 0 | 7 | 6 | -1.14 |
| M20 | 116 | 0.5 | -10.7 | 1 | 0 | 0 | 0 | 8 | 7 | -0.17 |
| M21 | 165 | 0.3 | -7.2 | 1 | 0 | 0 | 0 | 3 | 11 | -0.02 |
| M22 | 168 | 0.3 | -8.7 | 0 | 1 | 0 | 0 | 4 | 11 | 0.08 |
| M23 | 187 | 0.65 | -10.8 | 0 | 1 | 0 | 0 | 5 | 1 | 0.68 |
| M24 | 216 | 0.4 | -7 | 0 | 0 | 1 | 0 | 2 | 6 | 0.67 |
| M25 | 282 | 0.3 | -8.6 | 1 | 0 | 0 | 0 | 3 | 5 | 0.28 |
| M26 | 314 | 0.4 | -8.3 | 1 | 0 | 0 | 0 | 6 | 7 | 0.03 |
| M27 | 322 | 0.4 | -11.3 | 0 | 0 | 0 | 1 | 6 | 6 | -0.05 |
| M28 | 377 | 0.35 | -10.3 | 1 | 0 | 0 | 0 | 5 | 6 | 0.22 |
| M29 | 388 | 0.25 | -7.1 | 1 | 0 | 0 | 0 | 2 | 8 | 0.17 |
| M30 | 389 | 0.25 | -7 | 0 | 0 | 0 | 1 | 2 | 9 | 0.18 |
| M31 | 411 | 0.4 | -9.9 | 0 | 0 | 1 | 0 | 4 | 5 | -0.60 |
| M32 | 442 | 0.4 | -7.3 | 0 | 1 | 0 | 0 | 6 | 9 | -0.07 |
| M33 | 477 | 0.4 | -7.2 | 0 | 1 | 0 | 0 | 6 | 8 | -0.11 |
| M34 | 518 | 0.3 | -7 | 1 | 0 | 0 | 0 | 4 | 10 | 0.43 |
| M35 | 525 | 0.4 | -7.6 | 1 | 0 | 0 | 0 | 7 | 8 | 0.02 |
| M36 | 647 | 0.25 | -7.1 | 0 | 1 | 0 | 0 | 3 | 9 | 0.34 |
| M37 | 687 | 0.5 | -7.1 | 0 | 1 | 0 | 0 | 7 | 8 | 0.54 |
| M38 | 692 | 0.45 | -8.6 | 0 | 1 | 0 | 0 | 6 | 7 | 0.41 |

PDC1p-Sapphire

CATGCGACTGGGTGAGCATATGTTCCGCTGATGTGATGTGCAAGATAAACAAGCAAG
S1
GCAGAAACTAACTTCTTCTTCATGTAATAAACACACCCCGCGTTTATTTACCTATCTCT

AAACTTCAACACCTTATATCATAACTAATATTTCTTGAGATAAGCACACTGCACCCATA

CCTTCCTTAAAAACGTAGCTTCCAGTTTTTGGTGGTTCCGGCTTCCTTCCCGATTCCG
S2 S3 S4
CCCGCTAAACGCATATTTTTGTTGCCTGGTGGCATTTGCAAAATGCATAACCTATGCAT
S5 S6
TTAAAAGATTATGTATGCTCTTCTGACTTTTCGTGTGATGAGGCTCGTGGAAAAAATG
S7 S8
AATAATTTATGAATTTGAGAACAATTTTGTGTTGTTACGGTATTTTACTATGGAATAATC
S9 S10
AATCAATTGAGGATTTATGCAAATATCGTTTGAATATTTTTCCGACCCTTTGAGTACTT
S11
TTCTTCATAATTGCATAATATTGTCCGCTGCCCCTTTTTCTGTTAGACGGTGTCTTGATC
S12
TACTTGCTATCGTTCAACACCACCTTATTTTCTAACTATTTTTTTTTTAGCTCATTTGAAT

CAGCTTATGGTGATGGCACATTTTTGCATAAACCTAGCTGTCCTCGTTGAACATAGGA
S13 S14
AAAAAAAATATATAAACAAGGCTCTTTCACTCTCCTTGCAATCAGATTTGGGTTTGTTC
S15
CCTTTATTTTCATATTTCTTGTCATATTCCTTTCTCAATTATTATTTTCTACTCATAACCTC

ACGCAAAATAACACAGTCAAATCAATCAAA**ATG**TCTAAAGGTGAAGAATTATTCACTG
S16
GTGTTGTCCCAATTTTGGTTGAATTAGATGGTGATGTTAATGGTCACAAATTTTCTGTC
S17 S18
TCCGGTGAAGGTGAAGGTGATGCTACTTACGGTAAATTGACCTTAAAATTTATTTGTA
S19 S20 S21 S22

CTACTGGTAAATTGCCAGTTCCATGGCCAACCTTAGTCACTACTTTTTCTTATGGTGTT
S23 S24 S25
ATGGTTTTTGCTAGATACCCAGATCATATGAAACAACATGACTTTTTCAAGTCTGCCAT

GCCAGAAGGTTATGTTCAAGAAAGAACTATTTTTTTCAAAGATGACGGTAACTACAA
S26 S27
GACCAGAGCTGAAGTCAAGTTTGAAGGTGATACCTTAGTTAATAGAATCGAATTAAAA
S28 S29
GGTATTGATTTTAAAGAAGATGGTAACATTTTAGGTCACAAATTGGAATACAACTTTAA
S30 S31 S32
CTCTCACAATGTTTACATCATGGCTGACAAACAAAAGAATGGTATCAAAGCTAACTTC
S33 S34
AAAATTAGACACAACATTGAAGATGGTGGTGTTCAATTAGCTGACCATTATCAACAAA
S35
ATACTCCAATTGGTGATGGTCCAGTCTTGTTACCAGACAACCATTACTTATCCATTCAA
S36
TCTGCCTTATCCAAAGATCCAAACGAAAAGAGAGACCACATGGTCTTGTTAGAATTTG

TTACTGCTGCTGGTATTACCCATGGTATGGATGAATTGTACAAA
S37 S38

| No. | Location | GC% | Energy | AGG | TGG | CGG | GGG | #G | #A | Fold change (log2) |
|---|---|---|---|---|---|---|---|---|---|---|
| S1 | -745 | 0.35 | -8.4 | 1 | 0 | 0 | 0 | 4 | 10 | -0.35 |
| S2 | -596 | 0.35 | -10.1 | 0 | 1 | 0 | 0 | 3 | 6 | -0.02 |
| S3 | -593 | 0.45 | -11.2 | 0 | 1 | 0 | 0 | 5 | 3 | -0.29 |
| S4 | -587 | 0.45 | -10.7 | 0 | 0 | 1 | 0 | 5 | 1 | 0.15 |
| S5 | -541 | 0.35 | -8.3 | 0 | 1 | 0 | 0 | 3 | 5 | 0.03 |
| S6 | -538 | 0.45 | -10.4 | 0 | 1 | 0 | 0 | 5 | 2 | -0.78 |
| S7 | -468 | 0.4 | -8.3 | 1 | 0 | 0 | 0 | 5 | 2 | -1.39 |
| S8 | -461 | 0.5 | -9.6 | 0 | 1 | 0 | 0 | 7 | 2 | -1.29 |
| S9 | -413 | 0.25 | -11.6 | 0 | 0 | 1 | 0 | 4 | 6 | -1.42 |
| S10 | -400 | 0.3 | -8.5 | 0 | 1 | 0 | 0 | 4 | 4 | -0.40 |
| S11 | -381 | 0.25 | -7.1 | 1 | 0 | 0 | 0 | 3 | 9 | -0.08 |
| S12 | -283 | 0.45 | -9.2 | 0 | 0 | 1 | 0 | 3 | 2 | -0.18 |
| S13 | -202 | 0.35 | -10.1 | 0 | 1 | 0 | 0 | 3 | 6 | 1.13 |
| S14 | -196 | 0.35 | -10 | 0 | 1 | 0 | 0 | 5 | 5 | -0.02 |
| S15 | -103 | 0.4 | -9 | 0 | 1 | 0 | 0 | 2 | 5 | -0.02 |
| S16 | 27 | 0.3 | -12 | 0 | 1 | 0 | 0 | 4 | 9 | -2.35 |
| S17 | 57 | 0.3 | -7.8 | 0 | 1 | 0 | 0 | 4 | 6 | -2.40 |
| S18 | 69 | 0.3 | -7 | 0 | 1 | 0 | 0 | 6 | 7 | -1.60 |
| S19 | 90 | 0.4 | -8.4 | 0 | 0 | 1 | 0 | 3 | 4 | -1.40 |
| S20 | 96 | 0.4 | -9.5 | 1 | 0 | 0 | 0 | 4 | 4 | -2.35 |
| S21 | 102 | 0.55 | -10.8 | 1 | 0 | 0 | 0 | 7 | 3 | -1.45 |
| S22 | 117 | 0.45 | -8.8 | 0 | 0 | 1 | 0 | 7 | 5 | -2.60 |
| S23 | 150 | 0.15 | -10.4 | 0 | 1 | 0 | 0 | 1 | 7 | -1.45 |
| S24 | 169 | 0.45 | -12.9 | 0 | 1 | 0 | 0 | 4 | 5 | -0.97 |
| S25 | 198 | 0.25 | -8.2 | 0 | 1 | 0 | 0 | 1 | 4 | -0.71 |
| S26 | 270 | 0.5 | -9.9 | 1 | 0 | 0 | 0 | 4 | 5 | -2.47 |
| S27 | 309 | 0.2 | -8 | 0 | 0 | 1 | 0 | 2 | 7 | -0.82 |
| S28 | 345 | 0.4 | -10 | 1 | 0 | 0 | 0 | 6 | 7 | -1.29 |
| S29 | 378 | 0.2 | -7.1 | 1 | 0 | 0 | 0 | 3 | 10 | -0.76 |
| S30 | 399 | 0.25 | -7.9 | 0 | 1 | 0 | 0 | 5 | 8 | -0.40 |
| S31 | 411 | 0.25 | -10.8 | 1 | 0 | 0 | 0 | 4 | 9 | -0.65 |
| S32 | 422 | 0.25 | -8.1 | 0 | 1 | 0 | 0 | 2 | 8 | -0.40 |
| S33 | 458 | 0.35 | -7.1 | 0 | 1 | 0 | 0 | 1 | 6 | -0.82 |
| S34 | 477 | 0.35 | -7.6 | 0 | 1 | 0 | 0 | 4 | 11 | -1.40 |
| S35 | 519 | 0.3 | -7 | 0 | 1 | 0 | 0 | 3 | 10 | -0.06 |
| S36 | 564 | 0.25 | -8.1 | 0 | 1 | 0 | 0 | 0 | 10 | -2.35 |
| S37 | 681 | 0.35 | -8.9 | 0 | 1 | 0 | 0 | 4 | 4 | -2.17 |
| S38 | 693 | 0.5 | -14.5 | 0 | 1 | 0 | 0 | 4 | 4 | -2.40 |

PGK1p-Venus

TCAGGCATGAACGCATCACAGACAAAATCTTCTTGACAAACGTCACAATTGATCCCTC

CCCATCCGTTATCACAATGACAGGTGTCATTTTGTGCTCTTATGGGACGATCCTTATTA
V1V2
CCGCTTTCATCCGGTGATAGACCGCCACAGAGGGGCAGAGAGCAATCATCACCTGCA
V3 V4 V5
AACCCTTCTATACACTCACATCTACCAGTGTACGAATTGCATTCAGAAAACTGTTTGC

ATTCAAAAATAGGTAGCATACAATTAAAACATGGCGGGCACGTATCATTGCCCTTATCT
V6 V7
TGTGCAGTTAGACGCGAATTTTTCGAAGAAGTACCTTCAAAGAATGGGGTCTCATCTT
V8
GTTTTGCAAGTACCACTGAGCAGGATAATAATAGAAATGATAATATACTATAGTAGAGA
V9
TAACGTCGATGACTTCCCATACTGTAATTGCTTTTAGTTGTGTATTTTTAGTGTGCAAG

TTTCTGTAAATCGATTAATTTTTTTTTCTTTCCTCTTTTTATTAACCTTAATTTTTATTTTA

GATTCCTGACTTCAACTCAAGACGCACAGATATTATAACATCTGCACAATAGGCATTTG

CAAGAATTACTCGTGAGTAAGGAAAGAGTGAGGAACTATCGCATACCTGCATTTAAA

GATGCCGATTTGGGCGCGAATCCTTTATTTTGGCTTCACCCTCATACTATTATCAGGGC
V10 V11
CAGAAAAAGGAAGTGTTTCCCTCCTTCTTGAATTGATGTTACCCTCATAAAGCACGTG
V12
GCCTCTTATCGAGAAAGAAATTACCGTCGCTCGTGATTTGTTTGCAAAAAGAACAAA

ACTGAAAAAACCCAGACACGCTCGACTTCCTGTCTTCCTATTGATTGCAGCTTCCAAT

TTCGTCACACAACAGGGTCCTAGCGACGGCTCACAGGTTTTGTAACAAGCAATCGAA
V13 V14 V15

GGTTCTGGAATGGCGGGAAAGGGTTTAGTACCACATGCTATGATGCCCACTGTGATCT
V16 V17
CCAGAGCAAAGTTCGTTCGATCGTACTGTTACTCTCTCTCTTTCAAACAGAATTGTCC

GAATCGTGTGACAACAACAGCCTGTTCTCACACACTCTTTTCTTCTAACCAAGGGGG
V18 V19
TGGTTTAGTTTAGTAGAACCTCGTGAAACTTACATTTACATATATATAAACTTGCATAAA

TTGGTCAATGCAAGAAATACATATTTGGTCTTTTCTAATTCGTAGTTTTTCAAGTTCTTA
V20
GATGCTTTCTTTTTCTCTTTTTTACAGATCATCAAGGAAGTAATTATCTACTTTTTACAA
V21
CAAATATAAAACA**ATG**TCTAAAGGTGAAGAATTATTCACTGGTGTTGTCCCAATTTTG
V22 V23
GTTGAATTAGATGGTGATGTTAATGGTCACAAATTTTCTGTCTCCGGTGAAGGTGAA
V24 V25 V26 V27
GGTGATGCTACTTACGGTAAATTGACCTTAAAATTGATTTGTACTACTGGTAAATTGCC
V28
AGTTCCATGGCCAACCTTAGTCACTACTTTAGGTTATGGTTTGCAATGTTTTGCTAGAT
V29 V30
ACCCAGATCATATGAAACAACATGACTTTTTCAAGTCTGCCATGCCAGAAGGTTATGT
V31
TCAAGAAAGAACTATTTTTTTCAAAGATGACGGTAACTACAAGACCAGAGCTGAAGT
V32
CAAGTTTGAAGGTGATACCTTAGTTAATAGAATCGAATTAAAAGGTATTGATTTTAAAG
V33
AAGATGGTAACATTTTAGGTCACAAATTGGAATACAACTATAACTCTCACAATGTTTAC
V34
ATCACTGCTGACAAACAAAAGAATGGTATCAAAGCTAACTTCAAAATTAGACACAAC

ATTGAAGATGGTGGTGTTCAATTAGCTGACCATTATCAACAAAATACTCCAATTGGTG

ATGGTCCAGTCTTGTTACCAGACAACCATTACTTATCCTATCAATCTGCCTTATCCAAA

GATCCAAACGAAAAGAGAGACCACATGGTCTTGTTAGAATTTGTTACTGCTGCTGGTA

TTACCCATGGTATGGATGAATTGTACAAA

| No. | Location | GC% | Energy | AGG | TGG | CGG | GGG | #G | #A | Fold change (log2) |
|---|---|---|---|---|---|---|---|---|---|---|
| V1 | -1200 | 0.4 | -13.6 | 0 | 1 | 0 | 0 | 5 | 2 | 0.18 |
| V2 | -1199 | 0.35 | -13.9 | 0 | 0 | 0 | 1 | 4 | 2 | 0.42 |
| V3 | -1172 | 0.4 | -12.6 | 0 | 0 | 1 | 0 | 1 | 3 | 0.11 |
| V4 | -1152 | 0.6 | -12.3 | 0 | 0 | 0 | 1 | 6 | 6 | 0.07 |
| V5 | -1151 | 0.6 | -12.1 | 0 | 0 | 0 | 1 | 7 | 6 | 0.60 |
| V6 | -1037 | 0.3 | -7 | 0 | 1 | 0 | 0 | 3 | 10 | 0.05 |
| V7 | -1034 | 0.3 | -7.9 | 0 | 0 | 1 | 0 | 3 | 10 | 0.24 |
| V8 | -965 | 0.35 | -10.2 | 0 | 1 | 0 | 0 | 4 | 10 | 0.15 |
| V9 | -930 | 0.45 | -8.9 | 1 | 0 | 0 | 0 | 4 | 5 | 0.22 |
| V10 | -645 | 0.35 | -12.7 | 0 | 1 | 0 | 0 | 4 | 6 | 0.82 |
| V11 | -601 | 0.35 | -7.7 | 1 | 0 | 0 | 0 | 0 | 5 | 0.27 |
| V12 | -589 | 0.4 | -7 | 1 | 0 | 0 | 0 | 4 | 8 | 0.40 |
| V13 | -409 | 0.4 | -7 | 1 | 0 | 0 | 0 | 1 | 7 | 0.01 |
| V14 | -397 | 0.5 | -7 | 0 | 0 | 1 | 0 | 4 | 8 | 0.61 |
| V15 | -389 | 0.65 | -10.3 | 1 | 0 | 0 | 0 | 6 | 4 | 0.51 |
| V16 | -356 | 0.45 | -8.5 | 0 | 1 | 0 | 0 | 6 | 7 | 0.68 |
| V17 | -353 | 0.45 | -9.6 | 0 | 0 | 1 | 0 | 7 | 6 | 0.63 |
| V18 | -199 | 0.35 | -8.7 | 1 | 0 | 0 | 0 | 0 | 5 | 1.33 |
| V19 | -196 | 0.4 | -9 | 0 | 0 | 0 | 1 | 2 | 4 | 0.51 |
| V20 | -132 | 0.15 | -8.2 | 0 | 1 | 0 | 0 | 1 | 10 | 0.24 |
| V21 | -39 | 0.3 | -7.1 | 1 | 0 | 0 | 0 | 1 | 5 | 1.71 |
| V22 | 27 | 0.3 | -12 | 0 | 1 | 0 | 0 | 4 | 9 | -0.29 |
| V23 | 44 | 0.45 | -11 | 0 | 1 | 0 | 0 | 4 | 3 | 0.10 |
| V24 | 57 | 0.3 | -7.8 | 0 | 1 | 0 | 0 | 4 | 6 | -1.82 |
| V25 | 69 | 0.3 | -7 | 0 | 1 | 0 | 0 | 6 | 7 | -1.21 |
| V26 | 90 | 0.4 | -8.4 | 0 | 0 | 1 | 0 | 3 | 4 | -0.17 |
| V27 | 96 | 0.4 | -9.5 | 1 | 0 | 0 | 0 | 4 | 4 | 0.21 |
| V28 | 150 | 0.2 | -8.6 | 0 | 1 | 0 | 0 | 2 | 7 | -0.06 |
| V29 | 192 | 0.4 | -11.1 | 1 | 0 | 0 | 0 | 1 | 5 | -1.66 |
| V30 | 198 | 0.3 | -7 | 0 | 1 | 0 | 0 | 3 | 5 | -0.09 |
| V31 | 270 | 0.5 | -9.9 | 1 | 0 | 0 | 0 | 4 | 5 | -0.27 |
| V32 | 309 | 0.2 | -8 | 0 | 0 | 1 | 0 | 2 | 7 | 0.09 |
| V33 | 378 | 0.2 | -7.1 | 1 | 0 | 0 | 0 | 3 | 10 | -0.13 |
| V34 | 399 | 0.25 | -7.9 | 0 | 1 | 0 | 0 | 5 | 8 | -0.49 |

**Table S4.** Key parameters of guide RNAs used in validation experiments (Eno2p-tdTomato).

CGCTCAGCATCTGCTTCTTCCCAAAGATGAACGCGGCGTTATGTCACTAACGACGTGC
T1
ACCAACTTGCGGAAAGTGGAATCCCGTTCCAAAACTGGCATCCCTAATTGATACATCT
T2 T3 T4
ACACACCGCACGCCTTTTTCTGAAGCCCACTTTCGTGGACTTTGCCATATGCAAAAT
T5
TCATGAAGTGTGATACCAAGTCAGCATACACCTCACTAGGGTAGTTTCTTTGGTTGTA
T6 T7
TTGATCATTTGGTTCATCGTGGTTCATTAATTTTTTTCTCCATTGCTTTCTGGCTTTGAT
T8 T9 T10
CTTACTATCATTTGGATTTTTGTCGAAGGTTGTAGAATTGTATGTGACAAGTGGCACCA
T11 T12 T13
AGCATATATAAAAAAAAAAAGCATTATCTTCCTACCAGAGTTGATTGTTAAAAACGTAT

TTATAGCAAACGCAATTGTAATTAATTCTTATTTTGTATCTTTTCTTCCCTTGTCTCAATC

TTTTATTTTTATTTTATTTTTCTTTTCTTAGTTTCTTTCATAACACCAAGCAACTAATACT

ATAACATACAATAATA**ATG**GTGAGCAAGGGCGAGGAGGTCATCAAAGAGTTCATGCG

CTTCAAGGTGCGCATGGAGGGCTCCATGAACGGCCACGAGTTCGAGATCGAGGGCG
T14 T15 T16 T17 T18
AGGGCGAGGGCCGCCCCTACGAGGGCACCCAGACCGCCAAGCTGAAGGTGACCAA
T19T20 T21T22 T23T24 T25 T26
GGGCGGCCCCCTGCCCTTCGCCTGGGACATCCTGTCCCCCCAGTTCATGTACGGCTCC
T27 T28 T29 T30
AAGGCGTACGTGAAGCACCCCGCCGACATCCCCGATTACAAGAAGCTGTCCTTCCCC

GAGGGCTTCAAGTGGGAGCGCGTGATGAACTTCGAGGACGGCGGTCTGGTGACCGT
T31T32 T33 T34 T35 T36
GACCCAGGACTCCTCCCTGCAGGACGGCACGCTGATCTACAAGGTGAAGATGCGCGG
T37 T38 T39 T40
CACCAACTTCCCCCCCGACGGCCCCGTAATGCAGAAGAAGACCATGGGCTGGGAGG
T41 T42 T43 T44 T45
CCTCCACCGAGCGCCTGTACCCCCGCGACGGCGTGCTGAAGGGCGAGATCCACCAG
T46 T47 T48
GCCCTGAAGCTGAAGGACGGCGGCCACTACCTGGTGGAGTTCAAGACCATCTACATG
T49
GCCAAGAAGCCCGTGCAACTGCCCGGCTACTACTACGTGGACACCAAGCTGGACATC
T50
ACCTCCCACAACGAGGACTACACCATCGTGGAACAGTACGAGCGCTCCGAGGGCCG
T51 T52 T53
CCACCACCTGTTCCTGGGGCATGGCACCGGCAGCACCGGCAGCGGCAGCTCCGGCA
T54 T55 T56 T57 T58 T59
CCGCCTCCTCCGAGGACAACAACATGGCCGTCATCAAAGAGTTCATGCGCTTCAAGG
T60 T61
TGCGCATGGAGGGCTCCATGAACGGCCACGAGTTCGAGATCGAGGGCGAGGGCGAG
T62 T63 T64 T65 T66
GGCCGCCCCTACGAGGGCACCCAGACCGCCAAGCTGAAGGTGACCAAGGGCGGCCC

T67 T68 T69 T70T71 T72
CCTGCCCTTCGCCTGGGACATCCTGTCCCCCCAGTTCATGTACGGCTCCAAGGCGTAC
T73T74 T75 T76
GTGAAGCACCCCGCCGACATCCCCGATTACAAGAAGCTGTCCTTCCCCGAGGGCTTC
T77T78
AAGTGGGAGCGCGTGATGAACTTCGAGGACGGCGGTCTGGTGACCGTGACCCAGGA
T79T80 T81 T82 T83 T84 T85
CTCCTCCCTGCAGGACGGCACGCTGATCTACAAGGTGAAGATGCGCGGCACCAACTT
T86
CCCCCCCGACGGCCCCGTAATGCAGAAGAAGACCATGGGCTGGGAGGCCTCCACCG
T87 T88 T89
AGCGCCTGTACCCCCGCGACGGCGTGCTGAAGGGCGAGATCCACCAGGCCCTGAAG
T90T91 T92
CTGAAGGACGGCGGCCACTACCTGGTGGAGTTCAAGACCATCTACATGGCCAAGAAG
T93 T94 T95 T96
CCCGTGCAACTGCCCGGCTACTACTACGTGGACACCAAGCTGGACATCACCTCCCAC
T97
AACGAGGACTACACCATCGTGGAACAGTACGAGCGCTCCGAGGGCCGCCACCACCT
T98 T99
GTTCCTGTACGGCATGGACGAGCTGTACAAG

| No. | Location | GC% | Energy | AGG | TGG | CGG | GGG | #G | #A | Fold change (log2) |
|---|---|---|---|---|---|---|---|---|---|---|
| T1 | -516 | 0.45 | -7.9 | 0 | 0 | 1 | 0 | 3 | 6 | -1.81 |
| T2 | -482 | 0.5 | -8.6 | 0 | 0 | 1 | 0 | 4 | 6 | 0.59 |
| T3 | -475 | 0.55 | -9.8 | 0 | 1 | 0 | 0 | 6 | 6 | -1.79 |
| T4 | -456 | 0.5 | -11.7 | 0 | 1 | 0 | 0 | 4 | 6 | -1.85 |
| T5 | -397 | 0.45 | -14.3 | 0 | 1 | 0 | 0 | 3 | 3 | -1.86 |
| T6 | -338 | 0.45 | -7.9 | 1 | 0 | 0 | 0 | 2 | 7 | -1.67 |
| T7 | -325 | 0.45 | -12.8 | 0 | 1 | 0 | 0 | 4 | 3 | -1.53 |
| T8 | -308 | 0.3 | -11.2 | 0 | 1 | 0 | 0 | 4 | 3 | 0.10 |
| T9 | -298 | 0.35 | -7.8 | 0 | 1 | 0 | 0 | 4 | 4 | -0.97 |
| T10 | -266 | 0.3 | -7 | 0 | 1 | 0 | 0 | 1 | 1 | -1.12 |
| T11 | -244 | 0.3 | -10 | 0 | 1 | 0 | 0 | 2 | 4 | -0.15 |
| T12 | -230 | 0.3 | -8 | 1 | 0 | 0 | 0 | 4 | 4 | -1.94 |
| T13 | -205 | 0.35 | -8.4 | 0 | 1 | 0 | 0 | 6 | 7 | -1.72 |
| T14 | 56 | 0.6 | -15.8 | 0 | 1 | 0 | 0 | 6 | 4 | -1.41 |
| T15 | 59 | 0.65 | -14.4 | 1 | 0 | 0 | 0 | 8 | 3 | -1.62 |
| T16 | 72 | 0.6 | -14.9 | 0 | 0 | 1 | 0 | 7 | 5 | -1.43 |
| T17 | 92 | 0.65 | -9.7 | 1 | 0 | 0 | 0 | 7 | 4 | -2.29 |
| T18 | 93 | 0.6 | -9.3 | 0 | 0 | 0 | 1 | 7 | 5 | -1.36 |
| T19 | 98 | 0.65 | -12.7 | 1 | 0 | 0 | 0 | 9 | 4 | -1.94 |
| T20 | 99 | 0.6 | -10.6 | 0 | 0 | 0 | 1 | 9 | 5 | -1.90 |
| T21 | 104 | 0.75 | -10.8 | 1 | 0 | 0 | 0 | 11 | 4 | -2.27 |
| T22 | 105 | 0.7 | -8.7 | 0 | 0 | 0 | 1 | 11 | 5 | -1.79 |
| T23 | 119 | 0.85 | -16.2 | 1 | 0 | 0 | 0 | 9 | 2 | -2.11 |
| T24 | 120 | 0.8 | -12.8 | 0 | 0 | 0 | 1 | 8 | 3 | -1.98 |
| T25 | 143 | 0.65 | -8.3 | 1 | 0 | 0 | 0 | 4 | 6 | -2.41 |
| T26 | 152 | 0.6 | -11.2 | 1 | 0 | 0 | 0 | 6 | 6 | -2.15 |
| T27 | 153 | 0.55 | -10.2 | 0 | 0 | 0 | 1 | 6 | 7 | -1.74 |
| T28 | 156 | 0.55 | -9.5 | 0 | 0 | 1 | 0 | 8 | 7 | -1.25 |
| T29 | 176 | 0.8 | -10.9 | 0 | 0 | 0 | 1 | 4 | 0 | -1.65 |
| T30 | 204 | 0.55 | -7.4 | 0 | 0 | 1 | 0 | 3 | 3 | -2.16 |
| T31 | 269 | 0.6 | -9.8 | 1 | 0 | 0 | 0 | 4 | 4 | -2.20 |
| T32 | 270 | 0.55 | -10.1 | 0 | 0 | 0 | 1 | 4 | 5 | -2.19 |
| T33 | 280 | 0.65 | -11.1 | 0 | 1 | 0 | 0 | 5 | 3 | -1.77 |
| T34 | 302 | 0.6 | -8.8 | 1 | 0 | 0 | 0 | 8 | 4 | 0.10 |
| T35 | 306 | 0.55 | -9.3 | 0 | 0 | 1 | 0 | 7 | 5 | -0.13 |
| T36 | 314 | 0.65 | -11.8 | 0 | 1 | 0 | 0 | 8 | 4 | -2.07 |
| T37 | 344 | 0.7 | -11.5 | 1 | 0 | 0 | 0 | 4 | 3 | -1.39 |
| T38 | 348 | 0.65 | -14.4 | 0 | 0 | 1 | 0 | 5 | 4 | -0.44 |
| T39 | 365 | 0.6 | -8.9 | 1 | 0 | 0 | 0 | 6 | 5 | -0.28 |
| T40 | 378 | 0.45 | -8.7 | 0 | 0 | 1 | 0 | 6 | 7 | -0.12 |
| T41 | 399 | 0.7 | -8.6 | 0 | 0 | 1 | 0 | 3 | 4 | 0.33 |
| T42 | 425 | 0.45 | -7 | 0 | 1 | 0 | 0 | 5 | 9 | -0.42 |
| T43 | 426 | 0.4 | -9.7 | 0 | 0 | 0 | 1 | 5 | 9 | 0.68 |
| T44 | 430 | 0.55 | -7 | 0 | 1 | 0 | 0 | 7 | 7 | 0.70 |
| T45 | 434 | 0.6 | -9 | 1 | 0 | 0 | 0 | 9 | 6 | -0.60 |
| T46 | 465 | 0.75 | -8.9 | 0 | 0 | 1 | 0 | 6 | 3 | -1.75 |
| T47 | 476 | 0.8 | -11 | 1 | 0 | 0 | 0 | 7 | 2 | 0.68 |
| T48 | 491 | 0.65 | -10.4 | 1 | 0 | 0 | 0 | 7 | 5 | -0.65 |
| T49 | 548 | 0.45 | -7 | 0 | 1 | 0 | 0 | 4 | 7 | 0.24 |
| T50 | 599 | 0.55 | -7.8 | 0 | 1 | 0 | 0 | 4 | 6 | -1.09 |
| T51 | 620 | 0.6 | -7.9 | 1 | 0 | 0 | 0 | 3 | 6 | -0.55 |
| T52 | 635 | 0.55 | -8.2 | 0 | 1 | 0 | 0 | 4 | 7 | -0.77 |
| T53 | 657 | 0.6 | -8.4 | 0 | 0 | 0 | 1 | 6 | 6 | -0.34 |
| T54 | 679 | 0.7 | -10.1 | 0 | 0 | 0 | 1 | 4 | 2 | -2.56 |
| T55 | 684 | 0.65 | -10.3 | 0 | 1 | 0 | 0 | 5 | 3 | -2.05 |
| T56 | 690 | 0.65 | -9.5 | 0 | 0 | 1 | 0 | 7 | 2 | -2.33 |
| T57 | 699 | 0.75 | -9.7 | 0 | 0 | 1 | 0 | 8 | 4 | -2.55 |
| T58 | 705 | 0.8 | -12.5 | 0 | 0 | 1 | 0 | 8 | 4 | -2.74 |
| T59 | 714 | 0.75 | -12 | 0 | 0 | 1 | 0 | 7 | 4 | -2.49 |
| T60 | 731 | 0.8 | -11 | 1 | 0 | 0 | 0 | 4 | 1 | -2.14 |
| T61 | 743 | 0.55 | -9.7 | 0 | 1 | 0 | 0 | 3 | 7 | -2.25 |
| T62 | 782 | 0.6 | -15.8 | 0 | 1 | 0 | 0 | 6 | 4 | -2.48 |
| T63 | 786 | 0.6 | -9.6 | 0 | 0 | 0 | 1 | 7 | 4 | -2.37 |
| T64 | 818 | 0.65 | -9.7 | 1 | 0 | 0 | 0 | 7 | 4 | -2.37 |
| T65 | 824 | 0.65 | -12.7 | 1 | 0 | 0 | 0 | 9 | 4 | -2.10 |
| T66 | 830 | 0.75 | -10.8 | 1 | 0 | 0 | 0 | 11 | 4 | -1.65 |
| T67 | 845 | 0.85 | -16.2 | 1 | 0 | 0 | 0 | 9 | 2 | -0.04 |
| T68 | 846 | 0.8 | -12.8 | 0 | 0 | 0 | 1 | 8 | 3 | -0.17 |
| T69 | 869 | 0.65 | -8.3 | 1 | 0 | 0 | 0 | 4 | 6 | 0.03 |
| T70 | 878 | 0.6 | -11.2 | 1 | 0 | 0 | 0 | 6 | 6 | -0.30 |
| T71 | 879 | 0.55 | -10.2 | 0 | 0 | 0 | 1 | 6 | 7 | -0.98 |
| T72 | 882 | 0.55 | -9.5 | 0 | 0 | 1 | 0 | 8 | 7 | -0.08 |
| T73 | 901 | 0.85 | -10.6 | 0 | 1 | 0 | 0 | 5 | 0 | -0.07 |
| T74 | 902 | 0.8 | -10.9 | 0 | 0 | 0 | 1 | 4 | 0 | -1.79 |
| T75 | 930 | 0.55 | -7.4 | 0 | 0 | 1 | 0 | 3 | 3 | 0.18 |
| T76 | 938 | 0.55 | -7.4 | 1 | 0 | 0 | 0 | 4 | 4 | 0.29 |
| T77 | 995 | 0.6 | -9.8 | 1 | 0 | 0 | 0 | 4 | 4 | -1.08 |
| T78 | 996 | 0.55 | -10.1 | 0 | 0 | 0 | 1 | 4 | 5 | 0.01 |
| T79 | 1006 | 0.65 | -11.1 | 0 | 1 | 0 | 0 | 5 | 3 | 0.03 |
| T80 | 1007 | 0.6 | -9.4 | 0 | 0 | 0 | 1 | 5 | 3 | -0.37 |
| T81 | 1028 | 0.6 | -8.8 | 1 | 0 | 0 | 0 | 8 | 4 | -0.06 |
| T82 | 1032 | 0.55 | -9.3 | 0 | 0 | 1 | 0 | 7 | 5 | 0.16 |
| T83 | 1035 | 0.55 | -8.1 | 0 | 0 | 1 | 0 | 8 | 5 | 0.26 |
| T84 | 1040 | 0.65 | -11.8 | 0 | 1 | 0 | 0 | 8 | 4 | 0.10 |
| T85 | 1055 | 0.7 | -14.3 | 1 | 0 | 0 | 0 | 7 | 2 | -0.90 |
| T86 | 1074 | 0.65 | -14.4 | 0 | 0 | 1 | 0 | 5 | 4 | -0.36 |
| T87 | 1125 | 0.7 | -8.6 | 0 | 0 | 1 | 0 | 3 | 4 | -0.38 |
| T88 | 1157 | 0.55 | -7.9 | 0 | 0 | 0 | 1 | 7 | 7 | 0.30 |
| T89 | 1160 | 0.6 | -9 | 1 | 0 | 0 | 0 | 9 | 6 | 0.35 |
| T90 | 1202 | 0.8 | -11 | 1 | 0 | 0 | 0 | 7 | 2 | -0.29 |
| T91 | 1203 | 0.75 | -10.2 | 0 | 0 | 0 | 1 | 7 | 3 | -0.15 |
| T92 | 1217 | 0.65 | -10.4 | 1 | 0 | 0 | 0 | 7 | 5 | -0.20 |
| T93 | 1232 | 0.65 | -9.5 | 1 | 0 | 0 | 0 | 5 | 5 | -0.32 |
| T94 | 1236 | 0.6 | -12.1 | 0 | 0 | 1 | 0 | 7 | 6 | -0.28 |
| T95 | 1239 | 0.65 | -11.7 | 0 | 0 | 1 | 0 | 8 | 5 | 0.14 |
| T96 | 1250 | 0.7 | -10.8 | 0 | 1 | 0 | 0 | 7 | 5 | -0.01 |
| T97 | 1325 | 0.55 | -7.8 | 0 | 1 | 0 | 0 | 4 | 6 | -2.05 |
| T98 | 1346 | 0.6 | -7.9 | 1 | 0 | 0 | 0 | 3 | 6 | -1.95 |
| T99 | 1361 | 0.55 | -8.2 | 0 | 1 | 0 | 0 | 4 | 7 | -1.96 |

**Table S5.** Binary regression decision tree model.

Decision nodes: GC%, #G, [a]Loc, #A, ΔG, TGG, GGG, AGG, CGG

| Group | Value | Decision path |
|---|---|---|
| 1 | 0.115 | <6 → <4 → <-563 |
| 2 | -0.871 | <6 → <4 → ≥-563 |
| 3 | 2.122 | <6 → ≥ 4 → <-13.45 |
| 4 | -1.044 | <6 → ≥ 4 → ≥-13.45 → <-408 → ≥-528 |
| 5 | 0.821 | <6 → ≥ 4 → ≥-13.45 → <-408 → <-528 → <-12.5 |
| 6 | 0.597 | <6 → ≥ 4 → ≥-13.45 → <-408 → <-528 → ≥-12.5 → ≥7 |
| 7 | 0.018 | <6 → ≥ 4 → ≥-13.45 → <-408 → <-528 → ≥-12.5 → <7 → <-8.1 |
| 8 | 0.240 | <6 → ≥ 4 → ≥-13.45 → <-408 → <-528 → ≥-12.5 → <7 → ≥ -8.1 |
| 9 | 2.364 | <6 → ≥ 4 → ≥-13.45 → ≥-408 → <-404 |
| 10 | 0.053 | <6 → ≥ 4 → ≥-13.45 → ≥-408 → ≥-404 → <4 → ≥7 |
| 11 | 0.564 | <6 → ≥ 4 → ≥-13.45 → ≥-408 → ≥-404 → <4 → <7 → <-8.4 |
| 12 | 2.323 | <6 → ≥ 4 → ≥-13.45 → ≥-408 → ≥-404 → <4 → <7 → ≥ -8.4 |
| 13 | -0.755 | <6 → ≥ 4 → ≥-13.45 → ≥-408 → ≥-404 → ≥4 → ≥-8.45 |
| 14 | -0.401 | <6 → ≥ 4 → ≥-13.45 → ≥-408 → ≥-404 → ≥4 → <-8.45 → <-394 |
| 15 | 0.526 | <6 → ≥ 4 → ≥-13.45 → ≥-408 → ≥-404 → ≥4 → <-8.45 → ≥-394 |
| 16 | -3.382 | ≥ 6 → <-9.15 → <838 → <0.525 → <18 |
| 17 | 0.319 | ≥ 6 → <-9.15 → <838 → <0.525 → ≥ 18 → = 1 |
| 18 | 0.055 | ≥ 6 → <-9.15 → <838 → <0.525 → ≥ 18 → = 0 → <4 |
| 19 | -0.400 | ≥ 6 → <-9.15 → <838 → <0.525 → ≥ 18 → = 0 → ≥4 → ≥5 |
| 20 | -0.542 | ≥ 6 → <-9.15 → <838 → <0.525 → ≥ 18 → = 0 → ≥4 → <5 → <0.275 |
| 21 | -1.662 | ≥ 6 → <-9.15 → <838 → <0.525 → ≥ 18 → = 0 → ≥4 → <5 → ≥0.275 → <3 |
| 22 | -1.097 | ≥ 6 → <-9.15 → <838 → <0.525 → ≥ 18 → = 0 → ≥4 → <5 → ≥0.275 → ≥3 → <-13.7 |
| 23 | -0.606 | ≥ 6 → <-9.15 → <838 → <0.525 → ≥ 18 → = 0 → ≥4 → <5 → ≥0.275 → ≥3 → ≥-13.7 → <-12.5 |
| 24 | -0.925 | ≥ 6 → <-9.15 → <838 → <0.525 → ≥ 18 → = 0 → ≥4 → <5 → ≥0.275 → ≥3 → ≥-13.7 → ≥-12.5 → <341 |
| 25 | -0.596 | ≥ 6 → <-9.15 → <838 → <0.525 → ≥ 18 → = 0 → ≥4 → <5 → ≥0.275 → ≥3 → ≥-13.7 → ≥-12.5 → ≥ 341 |
| 26 | -0.134 | ≥ 6 → <-9.15 → <838 → ≥0.525 → <585 → ≥346 |
| 27 | -0.484 | ≥ 6 → <-9.15 → <838 → ≥0.525 → <585 → <346 → <2 |
| 28 | -0.911 | ≥ 6 → <-9.15 → <838 → ≥0.525 → <585 → <346 → ≥ 2 → ≥ 9.6 |
| 29 | -1.488 | ≥ 6 → <-9.15 → <838 → ≥0.525 → <585 → <346 → ≥ 2 → <9.6 → <82 |
| 30 | -1.706 | ≥ 6 → <-9.15 → <838 → ≥0.525 → <585 → <346 → ≥ 2 → <9.6 → ≥ 82 → < 4 |
| 31 | -2.083 | ≥ 6 → <-9.15 → <838 → ≥0.525 → <585 → <346 → ≥ 2 → <9.6 → ≥ 82 → ≥ 4 |
| 32 | -1.654 | ≥ 6 → <-9.15 → <838 → ≥0.525 → ≥585 → ≥827 |
| 33 | -2.290 | ≥ 6 → <-9.15 → <838 → ≥0.525 → ≥585 → <827 → =0 |
| 34 | -2.528 | ≥ 6 → <-9.15 → <838 → ≥0.525 → ≥585 → <827 → =1 |
| 35 | -0.953 | ≥ 6 → <-9.15 → ≥838 → <5 |
| 36 | -0.009 | ≥ 6 → <-9.15 → ≥838 → ≥5 → ≥8 |
| 37 | -0.980 | ≥ 6 → <-9.15 → ≥838 → ≥5 → <8 → ≥7 |
| 38 | -0.628 | ≥ 6 → <-9.15 → ≥838 → ≥5 → <8 → <7 → <-13.2 |
| 39 | 0.157 | ≥ 6 → <-9.15 → ≥838 → ≥5 → <8 → <7 → ≥-13.2 → ≥-9.4 |
| 40 | -0.273 | ≥ 6 → <-9.15 → ≥838 → ≥5 → <8 → <7 → ≥-13.2 → <-9.4 → =0 |
| 41 | -0.018 | ≥ 6 → <-9.15 → ≥838 → ≥5 → <8 → <7 → ≥-13.2 → <-9.4 → =1 |
| 42 | -1.983 | ≥ 6 → ≥-9.15 → ≥1243 |
| 43 | 0.028 | ≥ 6 → ≥-9.15 → <1243 → <147 → ≥8 |
| 44 | -2.099 | ≥ 6 → ≥-9.15 → <1243 → <147 → <8 → ≥0.575 |
| 45 | -0.604 | ≥ 6 → ≥-9.15 → <1243 → <147 → <8 → <0.575 → <5 |
| 46 | -1.196 | ≥ 6 → ≥-9.15 → <1243 → <147 → <8 → <0.575 → ≥5 |
| 47 | -1.242 | ≥ 6 → ≥-9.15 → <1243 → ≥147 → <4 |
| 48 | 0.193 | ≥ 6 → ≥-9.15 → <1243 → ≥147 → ≥4 → <1 |
| 49 | 0.324 | ≥ 6 → ≥-9.15 → <1243 → ≥147 → ≥4 → ≥1 → <-7.5 → =0 → ≥1141 |
| 50 | 0.036 | ≥ 6 → ≥-9.15 → <1243 → ≥147 → ≥4 → ≥1 → <-7.5 → =0 → <1141 → <6 |

**Cont. Table S5.** Binary regression decision tree model.

| Group | Value | Decision path |
|---|---|---|
| 51 | -0.206 | ≥ 6 → ≥-9.15 → <1243 → ≥147 → ≥4 → ≥1 → <-7.5 → =0 → <1141 → ≥6 → <8 |
| 52 | -0.605 | ≥ 6 → ≥-9.15 → <1243 → ≥147 → ≥4 → ≥1 → <-7.5 → =0 → <1141 → ≥6 → ≥8 |
| 53 | -0.209 | ≥ 6 → ≥-9.15 → <1243 → ≥147 → ≥4 → ≥1 → <-7.5 → =1 → <441 |
| 54 | -0.745 | ≥ 6 → ≥-9.15 → <1243 → ≥147 → ≥4 → ≥1 → <-7.5 → =1 → ≥441 |
| 55 | 0.511 | ≥ 6 → ≥-9.15 → <1243 → ≥147 → ≥4 → ≥1 → ≥-7.5 → ≥0.475 |
| 56 | 0.671 | ≥ 6 → ≥-9.15 → <1243 → ≥147 → ≥4 → ≥1 → ≥-7.5 → <0.475 → =1 |
| 57 | 0.158 | ≥ 6 → ≥-9.15 → <1243 → ≥147 → ≥4 → ≥1 → ≥-7.5 → <0.475 → =0 → ≥498 |
| 58 | 0.178 | ≥ 6 → ≥-9.15 → <1243 → ≥147 → ≥4 → ≥1 → ≥-7.5 → <0.475 → =0 → <498 → =1 |
| 59 | -0.067 | ≥ 6 → ≥-9.15 → <1243 → ≥147 → ≥4 → ≥1 → ≥-7.5 → <0.475 → =0 → <498 → =0 → <-7.15 |
| 60 | -0.252 | ≥ 6 → ≥-9.15 → <1243 → ≥147 → ≥4 → ≥1 → ≥-7.5 → <0.475 → =0 → <498 → =0 → ≥-7.15 |

**Table S6.** Measured and predicted concentrations of products from violacein pathway

| Strains | Measured concentrations (A.U.) | | | | Predicted concentrations (A.U.) | | | |
|---|---|---|---|---|---|---|---|---|
| | PV | PDV | V | DV | PV | PDV | V | DV |
| TEF1p(H) | 472.00 | 216.30 | 229.70 | 22.90 | 481.83 | 219.39 | 220.37 | 27.13 |
| TEF1p(M) | 392.70 | 206.80 | 294.00 | 32.30 | 422.47 | 191.06 | 191.92 | 24.63 |
| TEF1p(L) | 345.70 | 138.30 | 177.00 | 16.30 | 356.13 | 159.59 | 160.32 | 21.71 |
| PDC1p(H) | 408.30 | 188.60 | 222.50 | 23.50 | 835.91 | 390.75 | 392.48 | 40.50 |
| PDC1p(M) | 297.30 | 142.80 | 168.70 | 2.10 | 417.21 | 188.55 | 189.41 | 24.41 |
| PDC1p(L) | 55.10 | 40.70 | 60.70 | 0.10 | 85.92 | 35.23 | 35.40 | 7.36 |
| TPI1p(M) | 270.00 | 134.20 | 120.70 | 14.20 | 537.68 | 102.32 | 200.50 | 25.40 |
| TPI1p(L) | 342.20 | 120.00 | 132.90 | 15.90 | 239.27 | 400.73 | 200.50 | 25.40 |
| ENO2p(H) | 375.20 | 168.60 | 147.10 | 17.40 | 357.22 | 192.65 | 283.68 | 32.35 |
| ENO2p(M) | 337.80 | 156.20 | 199.20 | 19.90 | 425.71 | 198.31 | 215.19 | 26.69 |
| ENO2p(L) | 455.60 | 231.70 | 74.30 | 24.70 | 598.05 | 216.54 | 42.85 | 8.46 |
| ACD | 235.15 | 161.80 | 251.65 | 34.20 | 57.17 | 363.64 | 295.96 | 36.48 |
| CD | 219.75 | 149.10 | 245.45 | 28.10 | 57.17 | 297.88 | 295.96 | 36.48 |
| AD | 251.95 | 286.35 | 81.10 | 23.40 | 202.19 | 377.62 | 150.95 | 22.50 |

Note: PV: Proviolacein, PDV: Prodeoxyviolacein, V: Violacein, DV: Deoxyviolacein